\documentclass[prb,preprint]{revtex4}
\usepackage{graphicx}
\usepackage{amsmath}
\usepackage{color}
\newcommand{\beq}{\begin{equation}}
 \newcommand{\eeq}{\end{equation}}
 \newcommand{\beqa}{\begin{eqnarray}}
 \newcommand{\eeqa}{\end{eqnarray}}

 \newcommand{\bm}{\boldmath}

\newcommand{\pdfstyle}{
\newcommand{\Endrule}{\vskip 3pt\noindent\hrule width 8.6cm\vskip 3pt}
\newcommand{\Beginrule}{\vskip 3pt\noindent\hbox{%
\vbox{\hbox to 9cm{\hfill}}\vbox{\hrule width 9cm}} \vskip 3pt}
\newcommand{\StartTwoColumn}{\begin{multicols}{2}}
\newcommand{\EndTwoColumn}{\end{multicols}}
\renewcommand{\narrowtext}{\Beginrule\begin{multicols}{2}}
\renewcommand{\widetext}{\end{multicols}\Endrule}
\newcommand{\NP}{}}
\newcommand{\D}{W}

\makeatletter
\newcommand\figcaption{\def\@captype{figure}\caption} 
\newcommand\tabcaption{\def\@captype{table}\caption}
\makeatother

 \def\dabl#1#2{\frac{{\rm d}{#1}}{{\rm d}{#2}}}
\begin{document}

\title{Microscopic Dynamics of thin hard rods}
\author{Matthias Otto, Timo Aspelmeier and Annette Zippelius}
\affiliation{Institut f\"ur Theoretische Physik, Friedrich-Hund-Platz 1, 
D-37077 G\"ottingen, Germany}

\begin{abstract}
	We analyze the microscopic dynamics and transport properties of a gas of thin hard rods. Based on the collision rules for hard needles we derive a
  hydrodynamic equation that determines the coupled translational and
  rotational dynamics of a tagged thin rod in an ensemble of identical
  rods.  Specifically, based on a Pseudo-Liouville operator for binary
  collisions between rods, the Mori-Zwanzig projection formalism is
  used to derive a continued fraction representation for the
  correlation function of the tagged particle's density, specifying
  its position and orientation. Truncation of the continued fraction
  gives rise to a generalised Enskog equation, which can be compared
  to the phenomenological Perrin equation for anisotropic diffusion.
  Only for sufficiently large density do we observe anisotropic
  diffusion, as indicated by an anisotropic mean square displacement,
  growing linearly with time.  For lower densities, the Perrin
  equation is shown to be an insufficient hydrodynamic description for
  hard needles interacting via binary collisions.  We compare our
  results to simulations and find excellent quantitative agreement for
  low densities and qualtitative agreement for higher densities.

%\pacs{PACS numbers: }
\end{abstract}

\maketitle

\section{Introduction}
Theoretical work on the transport behavior of non-spherical molecules was
pioneered by the work of Debye \cite{debye:29} and Perrin \cite{perrin:34,perrin:36}.
Brownian motion of a rotational ellipsoid in a viscous solvent is described in
terms of a partial differential equation coupling rotational and translational
diffusion (see also Ref.~\onlinecite{berne:76}).  The central quantity is $\rho({\bf
  r},{\bf u},t;{\bf r}',{\bf u}',0)$, the probability to find a molecule with
center of mass (CMS) position ${\bf r}$ and orientation ${\bf u}$ at time $t$,
given that its CMS position is ${\bf r}'$ and its orientation is ${\bf u}'$ at
time $t=0$.  Its dynamical evolution is governed by the following
translational-rotational diffusion equation, first introduced by Perrin 
\cite{perrin:34,perrin:36}:
\begin{align}
\label{DE}
\partial_t \rho({\bf r},{\bf u},t;{\bf r}',{\bf u}',0)
&= \left[- D_R\hat{L}^2
 + D_\parallel(\nabla_{\bf r}\cdot{\bf u} )^2 + D_\perp
\left(\nabla_{\bf r}^2-(\nabla_{\bf r}\cdot{\bf u})^2\right) \right]
\rho({\bf r},{\bf u},t;{\bf r}',{\bf u}',0).
\end{align}
Here ${\bf u}$ denotes the orientation of the symmetry axis of the
molecule. The operator $\hat{L}^2$ is the familiar angular momentum
operator. It describes rotational diffusion, $D_R$ being the
rotational diffusion constant.  The usual Laplacian has been
decomposed into directions parallel and perpendicular to the particle
orientation, with parallel and perpendicular diffusion constants
$D_\parallel$ and $D_\perp$ respectively.

Transport coefficients may be calculated using the Kirkwood-Riseman theory
\cite{riseman} for rod-like molecules in solution. Broersma \cite{broersma}
computes the rotational and isotropic translational diffusion coefficients
approximately for long cylinders with no stick boundary conditions. Rods with
capped hemispheres at the ends have also been considered
\cite{yoshizaki,schilling}. Alternatively the rods are composed of spherical
hydrodynamic subunits \cite{tirado:79,tirado:84,eimer} with each sphere
interacting with the other spheres via hydrodynamic interactions. The
resulting transport coefficients obey the Stokes-Einstein relation, i.e. both
the rotational and the translational diffusion constant are inversely
proportional to the solvent viscosity. Transport properties of nonspherical molecules have also been computed using kinetic theory \cite{Cole:1984a,Cole:1985,Cole:1986,Evans:1988}.

In polymer physics, the transport behavior of rod-like macromolecules
has been investigated where anisotropic translational diffusion and
the coupling of rotational and translational diffusion is most
pronounced. The coupling of rotational and translational diffusion can
be detected experimentally by depolarized light scattering.  Dynamic
light scattering from cylindrically symmetric rigid particles has been
discussed by Aragon and Pecora \cite{aragon:77,aragon:80,aragon:85}. 
If only autocorrelations
are included, the intensity of scattered light is proportional to the
intermediate scattering function
\begin{eqnarray}
S({\bf k},t)&=&\frac{1}{N}\sum_{i=1}^N \langle
\alpha^{(i)}(0)\alpha^{(i)}(t) e^{i{\bf k}
({\bf r}^i(t)-{\bf r}^i(0))}\rangle \nonumber\\
\mbox{with} \quad \alpha^{(i)} &=& \sum_{\nu,\mu=1}^3
n_{\nu}^{\text{inc}}\alpha_{\nu \mu}^{(i)} 
 n_{\mu}^{\text{fin}}.\nonumber
\end{eqnarray}
Here $\mathbf{k}$ is the wave vector and $\alpha_{\nu \mu}^{(i)}$ is the polarizability tensor of particle
$i$ and ${\bf n}^{\text{inc}}({\bf n}^{\text{fin}})$ denotes the polarization vector
of the incident (scattered) light. Aragon and Pecora have solved the
Perrin equation, allowing for a comparison of the theoretical
predictions of the Perrin equation and the measured intensities.  If
the polarization of the incident and scattered light are orthogonal,
then the scattering intensity is proportional to
\begin{equation}
S({\bf k},t) \sim \frac 1N \sum_{i=1}^N
\langle({\bf u}_i(0)\cdot{\bf u}_i(t))^2 \, e^{i{\bf
    k}({\bf r}^i(t)-{\bf r}^i(0))}\rangle
\end{equation}
If the average over orientation and position is factorized, the
intermediate scattering function decay with a rate $\Gamma=6 D_R
+D_{\text{iso}} k^2$. The factorizing assumption, however, is valid only
if the respective decay rates for orientational und positional degrees of freedom 
differ sufficiently. This is usually the case as long as $D_R\gg D_{\text{iso}} k^2$, which
is guaranteed for small wave vectors. Hence translational and rotational diffusion
coefficients can and have been obtained from polarized light scattering
experiments \cite{zero,lehner,cush}, however the experiments are
difficult due to low intensities of the scattered light
\cite{russo:93}. Since rotational and lateral 
motions of a rod-like molecule are usually studied in light scattering experiments for 
$kL\ge 1$, where $L$ is the rod length \cite{russo:93}, the factorizing assumption is bound to break down and coupling of translational and rotational diffusion will become important.

Even though the Perrin equation was originally formulated for the Brownian
motion of a rodlike macromolecule in solution, it is usually thought to be
applicable to the motion of a tagged rod in a fluid of otherwise identical
rods. These systems have been simulated in the limit that the diameter of the
rods goes to zero. Frenkel and Maguire \cite{frenkel:83} suggested an
algorithm for the dynamics of hard needles and furthermore computed the
autocorrelation of the linear velocity and the angular velocity in the Enskog
approximation. The latter breaks down when the density $\rho_0$ is
sufficiently high, such that the tagged rod cannot rotate unless other rods
move out of the way. This so called semidilute transition density was
estimated to be \cite{magda:86} $\rho_0=\frac{N}{V}L^3\sim 70$. Here $N$
denotes the total number of rods of length $L$ confined to a volume $V$. For
larger densities the rotational diffusion constant $D_R\sim \rho_0^{-2}$, as
first suggested by Doi and Edwards with help of a scaling argument
\cite{edwards:86}. They argue that rotational diffusion of a rod can be
described by a random walk, so that the rotational diffusion constant is given
by the product of the jump frequency, $\nu$, and the average step size
squared: $D_R=\nu (a/L)^2$.  In the semidilute regime the jump
frequency is $\nu\sim D_{\parallel}/L^2$ and the step size is $(a/L)\sim
1/\rho_0$, leading to the above scaling of $D_R$ with density. A similar
scaling argument has been formulated for the transverse diffusion coefficient.
The inverse of the jump frequency $\nu$ can also be interpreted as the
disentanglement time, or within the reptation model as the time needed for the
breakup of the tube, confining the test rod. On time scales of order $1/\nu$
the rod can diffuse in the transverse direction by an amount $a$, giving rise
to $D_{\perp}\sim a^2 \nu \sim \rho_0^{-2}$. Such a scaling has been derived
by Szamel and Schweizer \cite{szamel:93,szamel:94} within a dynamic mean field
theory for self and tracer diffusion. They neglect rotational motion of the
rods and close the hierarchy of dynamical distribution functions on the two
particle level.

In this work, we discuss the microscopic dynamics of a gas of
infinitely thin needles as a minimal model for rod-like
macromolecules. An approximate kinetic theory is derived which allows
us to determine the time delayed autocorrelation function of a tagged
particle's density with a given orientation ${\bf u}$.  We derive a
hydrodynamic equation for this autocorrelation function from the
microscopic collisional dynamics and examine the validity of the
phenomenological Perrin equation in the limit of small wave vectors
$kL\ll 1$. In fact, differences with the Perrin theory are found. In
particular, meaningful anisotropic diffusion constants $D_{\parallel}$
and $D_{\perp}$ can only be obtained in the limit of high densities.
In contrast to the Perrin theory we include the free motion of the
particles in the microscopic theory, giving rise to ballistic behavior at short times. In addition to smooth needles, we consider
rough needles \cite{chapman}, which allow for a change of the relative
tangential velocity in a collision.  The results, however, are
qualitatively the same as for smooth needles.

The outline of the paper is as follows. In Sec.~\ref{sec:perrin} we
analyse the solutions of the Perrin equation following Aragon and
Pecora. In particular we discuss anisotropic diffusion with help of
the mean square displacement parallel and pependicular to the rod's
orientation. In Sec.~\ref{sec:coll} we analyse binary collisions of
two hard needles and introduce a Pseudo Liouville operator for the
time evolution of a gas of hard needles. Subsequently
(Sec.~\ref{sec:proj}) we apply the projection operator formalism of
Mori and Zwanzig to the correlation function of interest, which is
thereby represented as a continued fraction. The formalism is applied
to the density of a tagged particle with given orientation, i.e.\
$\rho({\bf r},{\bf u})$, in Sec.~\ref{density}. In Sec.~\ref{sec:num} 
we describe numerical simulations. The results of a 
truncation of the continued fraction are presented and compared with the simulation results in Sec.~\ref{sec:results}. 
We conclude with a discussion and outlook in
Sec.~\ref{sec:diss}.

\section{Analysis of the Perrin Equation}
\label{sec:perrin}
In this section we review the predictions of the Perrin equation,
Eq.~\eqref{DE}, as discussed in Refs.~\onlinecite{aragon:77,aragon:80,aragon:85}, with particular
emphasis on anisotropic diffusion. We do not present any new results,
instead we want to familiarize the reader with this classic
phenomenological theory.  Later on, we will compare this theory with
our kinetic theory, which is based on a microscopic dynamic model.

The first step in the solution of the Perrin equation is a Fourier
transform with respect to the spatial coordinates
\begin{equation}
\tilde{\rho}({\bf k},{\bf u},{\bf u}';t)=
\int d^3 r\; e^{i{\bf k}{\bf r}}
\rho({\bf r},{\bf u},t;{\bf r}',{\bf u}',0)
\end{equation}
where we have assumed spatial homogeneity.
The Fourier transformed function is the solution of
\begin{align}
\label{DEK}
\partial_t 
\tilde{\rho}({\bf k},{\bf u},{\bf u}',t)
&= [-D_R\hat{L}^2  -k^2 D_{\perp}- 
(D_\parallel-D_{\perp}) 
({\bf k}\cdot {\bf u})^2]
\tilde{\rho}({\bf k},{\bf u},{\bf u}',t).
\end{align}
It is convenient to choose a coordinate frame ${\bf k}=k\hat{{\bf z}}$
with $k=|{\bf k}|$,
so that the unit vector ${\bf u}$ is specified by a polar angle $\phi$ and 
$\eta={\bf
  k}\cdot{\bf u}/(k)$. In terms of $\phi$ and $\eta$
the angular momentum operator is given explicitly by: 
\begin{equation}
\label{angular_momentum}
\hat{L}^2=-\frac{\partial}{\partial \eta}(1-\eta^2)
\frac{\partial}{\partial\eta}-
\frac{1}{1-\eta^2}\frac{\partial^2}{\partial \phi^2}
\end{equation}
Provided $\gamma^2:=(D_{\parallel}-D_{\perp})k^2/D_R > 0$, the right
hand side of Eq.~(\ref{DEK}) can be identified with the anisotropic
Laplacian, whose eigenfunctions, 
$S_{m,l}(\gamma,\eta)e^{im\phi}$, are known and expressed in terms of the
prolate spheroidal wavefunctions $S_{lm}$ \cite{flammer:57,stratton}.
The latter solve the equation
\begin{align}
\left(-\frac{d}{d\eta}(1-\eta^2)\frac{d}{d\eta}
  +\frac{m^2}{1-\eta^2}+\gamma^2\eta^2 \right)S_{lm}(\gamma,\eta)
&= \lambda_{lm}(\gamma^2) S_{lm}(\gamma,\eta) 
\end{align}
with eigenvalues
\begin{equation}
\lambda_{lm}(\gamma^2)=l(l+1)+\frac{2l(l+1)-2m^2-1}{(2l-1)(2l+3)}\gamma^2
+{\cal O} (\gamma^4) 
\end{equation}
We are interested in diffusion and hence consider only long wavelength
phenomena, so that it suffices to keep terms up to and including
${\cal O}(\gamma^2)$.

The full solution of the Perrin equation can then be written in terms
of the eigenfunctions
\begin{align}
\label{Aragon}
\tilde{\rho}({\bf k},{\bf u},{\bf u}',t) 
&=\frac{1}{8\pi^2}\sum_{m=0}^{\infty}
\sum_{l=m}S_{lm}(\gamma,\eta) S_{lm}(\gamma,\eta') e^{im\phi}
e^{-im\phi'} 
e^{-\Gamma_{lm}(\gamma^2)t}
\end{align}
such that the eigenvalues determine the relaxation rates according to
$\Gamma_{lm}(\gamma^2)=D_{\perp} k^2 +D_R \lambda_{lm}(\gamma^2)$.

The anisotropic diffusive behaviour can be analysed and has been discussed
in the literature in terms of the mean square
displacement. We define the displacement $\Delta {\bf r}(t):=
{\bf r}_0(t)-{\bf r}_0(0)$ and compute its projection on the initial
orientation $\Delta r_{\parallel}(t):=\Delta {\bf r}(t)\cdot{\bf
  u}_0(0)$ 
as well as perpendicular
to it $\Delta r_{\perp}(t):=\Delta {\bf r}(t)-{\bf u}_0(0) \Delta {\bf r}(t)
\cdot{\bf u}_0(0)$. 
The Perrin equation predicts
\begin{eqnarray}
\label{displacementpar}
\langle(\Delta r_{\parallel}(t))^2\rangle&=& 2 D_{\text{iso}}t+
\frac{2(D_{\parallel}-D_{\perp})}{9D_R}\left(1-e^{-6 D_R t}\right)\\
\label{displacementperp}
\langle(\Delta r_{\perp}(t))^2\rangle&=& 4 D_{\text{iso}}t-
\frac{2(D_{\parallel}-D_{\perp})}{9D_R}\left(1-e^{-6 D_R t}\right).
\end{eqnarray}
Analysing the short time behaviour one finds {\it anisotropic}
diffusion, 
$\langle(\Delta r_{\parallel}(t))^2\rangle  \sim  2D_{\parallel} t$
and $\langle(\Delta r_{\perp}(t))^2\rangle  \sim  4D_{\perp} t$,
such that in general the diffusive motion parallel to the initial
orientation is faster than perpendicular to it.
The asymptotic long time behaviour is again diffusive with, however,
an {\it isotropic} diffusion constant
$D_{\text{iso}}=(D_{\parallel}+2D_{\perp})/3$.
The crossover is determined by the rotational diffusion constant
$D_R$, such that for $6D_R t \gg 1$ the initial orientation has been
forgotten and diffusion is completely isotropic.

The above discussion already indicates that it might be difficult to
detect anisotropic diffusion, since it is not the true asymptotic long
time behaviour and, on the other hand, we expect ballistic motion for short times which is not accounted for in the Perrin equation. Hence the
anisotropic diffusion on intermediate timescales will be superimposed
by a crossover to ballistic motion on short time scales and isotropic
diffusion on long time scales. Only if the density is sufficiently
high, so that the time for rotation of the molecule is large, do we
expect to see a clear separation of the two diffusive regimes. 

Several attempts have been made to detect anisotropic diffusion from
velocity autocorrelations. For isotropic diffusion there is a
simple relation, connecting the asymptotic behaviour of the mean
square displacement with the velocity autocorrelation, integrated over time.
For anisotropic diffusion $\langle(\Delta r_{\parallel}(t))^2\rangle$ is
\textit{not} simply related to the autocorrelation of $v_{\parallel}(t)={\bf
  v}(t)\cdot {\bf u}(0)$, as noted in Ref.~\onlinecite{magda:86}.

\section{Dynamics of hard needles}
\label{sec:coll}
\subsection{Collision rules}
%general
%perfectly smooth
%perfectly rough
The phase space of needles is specified by the needles' translational and
angular 
velocities ${\bf v}_i$, $\hbox{\bm $\omega$}_i$ and their center of mass
positions ${\bf r}_i$ and orientations ${\bf u}_i$, with $i=0\dots N$.
The needles have length $L$, mass $m$ and moment of
inertia $I$. Given two needles, labeled $0$ and $1$, their
orientations define a
plane $E_{01}$ whose normal vector is given by
(see Fig.~\ref{2rods}):
\beq
{\bf u}_\perp=\frac{{\bf u}_0\times{\bf u}_1}{|{\bf u}_0\times{\bf u}_1|}
\eeq
The three vectors ${\bf u}_0$, ${\bf u}_\perp$ and
\beq
{\bf u}_0^\perp=\frac{{\bf u}_1-({\bf u}_0\cdot{\bf u}_1){\bf u}_0}{\sqrt{1-({\bf u}_0\cdot{\bf u}_1)^2}},
\eeq
define a local orthonormal frame which is convenient for 
our calculations. For ${\bf u}_1^\perp$ there is an analoguous expression
if one permutes indices $0$ and $1$ in the last equation.
Let ${\bf r}_{01}={\bf r}_0-{\bf r}_1$ be the CMS
distance between two needles, which may be decomposed into a
component perpendicular ${\bf r}_{01}^\perp=({\bf r}_{01}\cdot {\bf
  u}_\perp){\bf u}_\perp$ and parallel ${\bf
  r}_{01}^\parallel=s_{01}{\bf u}_0-s_{10}{\bf u}_1$ to $E_{01}$ \cite{cichocki:87}.
The parameters $s_{ij}$
are determind from 
\beq
s_{ij}=-\frac{{\bf r}_{ij}\cdot{\bf u}_j^\perp}{({\bf u}_i\cdot{\bf u}_j^\perp)}
\eeq
Note that ${\bf u}_1^\perp=-\sqrt{1-({\bf u}_0\cdot{\bf u}_1)^2}{\bf
  u}_0+({\bf u}_0\cdot{\bf u}_1){\bf u}^\perp_0$.

Let us recall the general collision rules for hard needles (see e.g.\
Ref.~\onlinecite{aspelmeier:01}, with the 
coefficient of the normal restitution $\epsilon=1$ and the coefficient
of tangential restitution $\beta=\pm 1$). The
relative velocity of the contact points is given by ${\bf V}={\bf
  v}_0-{\bf v}_1+s_{01}\dot{\bf u}_0-s_{10}\dot{\bf u}_1$ and 
can be
 decomposed with respect to 
the orthonormal frame $({\bf u}_\perp,{\bf u}_0^\perp,{\bf u}_0)$
according to
${\bf V} = x{\bf u}_\perp + y{\bf u}^\perp_0 + z{\bf u}_0$.
The normal and tangential component of the relative velocity after the
collision
(primed quantities) are given in terms of the respective components
before the collision as follows:
\beqa
({\bf u}_\perp\cdot {\bf V}^\prime)&=&
-({\bf u}_\perp\cdot {\bf V}) \\
({\bf u}_\perp\times {\bf V}^\prime)&=&
-\beta({\bf u}_\perp\times {\bf V})
\eeqa
Here $\beta=-1$ for smooth needles, and $\beta=1$ for totally rough
needles, being the only values allowed by energy conservation.
In terms of the particle's translational and angular velocities the
collision rules are the following:
\begin{align}
\begin{split}
	\label{coll.rules}
	{\bf p}^\prime_0 &= {\bf p}_0+\Delta{\bf p}\\
	{\bf p}^\prime_1 &= {\bf p}_1-\Delta{\bf p}\\
	\hbox{\bm $\omega$}^\prime_0 &=
	  \hbox{\bm $\omega$}_0+\frac{s_{01}}{I}{\bf u}_0\times\Delta{\bf p}\\
	\hbox{\bm $\omega$}^\prime_1 &=
	  \hbox{\bm $\omega$}_1-\frac{s_{10}}{I}{\bf u}_1\times\Delta{\bf p}.
\end{split}
\end{align}

\subsection{Perfectly smooth needles: $\beta=-1$ }
In this case, the translational momentum transfer occurs only
perpendicular to the plane $E_{01}$, and is given by
$\Delta{\bf p}=\alpha {\bf u}_\perp$
with
\beq
\label{alpha}
\alpha=-m
\frac{({\bf u}_\perp\cdot {\bf V})}{1+\left(\frac{m}{2I}(s_{01}^2+s_{10}^2)\right)}
\eeq
The temporal change of orientation $\dot{\bf u}_0$ is changed by a 
collision due to translational momentum transfer,
$I\, \Delta\dot{\bf u}_0 = s_{01}\Delta{\bf p}$.

\subsection{Perfectly rough needles: $\beta=1$ }
Rough needles also allow for a momentum transfer in
the directions ${\bf u}_0$ and ${\bf u}_0^\perp $:
\beq
\label{p.rough}
\Delta{\bf p}=\alpha {\bf u}_\perp+\gamma_1{\bf u}_0+\gamma_2{\bf u}_0^\perp. 
\eeq
The coefficients
 $\gamma_1$ and $\gamma_2$ are given by:
\beqa
\label{gamma1u2}
\gamma_1 &=& \frac{1+\beta}{2} \frac{By-Cz}{AC-B^2}\\
\gamma_2 &=& \frac{1+\beta}{2} \frac{Bz-Ay}{AC-B^2}
\eeqa
with
\beqa
A &=& \frac{1}{m}+\frac{s_{10}^2}{2I}(1-({\bf u}_0\cdot{\bf u}_1)^2)\\
B &=& -\frac{s_{10}^2}{2I}({\bf u}_0\cdot{\bf u}_1)\sqrt{1-({\bf
    u}_0\cdot{\bf u}_1)^2}\\
C &=& \frac{1}{m}+\frac{s_{01}^2}{2I}+\frac{s_{10}^2}{2I}({\bf u}_0\cdot{\bf u}_1)^2
\eeqa
and the variables $y,z$ as defined above.
The change of angular velocity is given by the expression:
\begin{equation}
\label{udot.rough}
\Delta\dot{\bf u}_0 = \frac{s_{01}}{I}\left(
\alpha {\bf u}_\perp +\gamma_2 {\bf u}_0^\perp
\right).
\end{equation}

\subsection{Liouville operator for hard needles}
The time evolution of any function $A(\Gamma, t)$ on phase space
$\Gamma=\{{\bf r}_i, {\bf u}_i, {\bf v}_i, \hbox{\bm $\omega$}_i \}$
is determined by the pseudo-Liouville operator ${\cal L}_+$
\cite{Cole:1985,Cole:1986}
\beq
A(\Gamma, t)=
\exp(i{\cal L}_+ t)A(\Gamma, 0), \; t>0
\eeq
The pseudo-Liouville operator ${\cal L}_+$ consists of two parts,
${\cal L}_+={\cal L}^0 +{\cal L}^\prime_+$.
The first part results from the Poisson bracket with the Hamiltonian 
of free translational motion and rotation,
$i{\cal L}^0=\{{\cal H},\dots\}$.
For $N$ needles the Hamiltonian expressed in terms of generalized
canonical momenta is given as follows:
\beq
{\cal H}=\sum_{i=1}^N\left(
\frac{1}{2m}{\bf p}_{{\bf r}_i}^2+\frac{1}{2I}p_{\theta_i}^2+
\frac{1}{2I\sin^2 (\theta_i)}p_{\phi_i}^2
\right)
\eeq
The second part of the pseudo-Liouville-operator, describing 
binary collisions, reads
\begin{align}
i{\cal L}^\prime_+
=\frac{1}{2}\sum_{i\neq j}
\left|\dabl{}{t}|{\bf r}_{ij}^\perp|\right|
\theta\left(-\dabl{}{t}|{\bf r}_{ij}^\perp|\right)
\theta\left(\frac{L}{2}-|s_{ij}|\right)
\theta\left(\frac{L}{2}-|s_{ji}|\right)
\delta\left(|{\bf r}_{ij}^\perp|-0^+\right)
\left(b_{ij}^+-1\right).
\end{align}
Here $\theta(x)$ is the Heaviside step function. The factor
$\left|\dabl{}{t}|{\bf r}_{ij}^\perp|\right|$
accounts for the flux of approaching needles (therefore requiring the factor 
$\theta\left(-\dabl{}{t}|{\bf r}_{ij}^\perp|\right)$). The
following two step functions and the delta functions enforce the
conditions $\frac{L}{2}>|s_{ij}|$, $\frac{L}{2}>|s_{ji}|$, and $|{\bf r}_{ij}^\perp|=0^+$ for a collision, assuming no
lateral extension of the needles. Finally $b_{ij}$, acting on a phase
space function $A(\Gamma)$, replaces linear and angular momenta before
a collision by those after collision according to the rules given in Eq.~(\ref{coll.rules}). One can equally consider the evolution backwards in time. The corresponding Liouville operator is given in Ref.~\onlinecite{aspelmeier:01}.

\section{Projection operator formalism}
\label{sec:proj}
In the following, autocorrelation functions of phase space functions
or observables $A(\Gamma,t)$ will be calculated: \beqa
\Phi(t)&=&\langle A^*(\Gamma,0)A(\Gamma,t)\rangle\nonumber\\
&=&\langle A^*(\Gamma,0)\exp(i{\cal L}_+ t)A(\Gamma,0)\rangle \eeqa
The bracket $\langle\dots\rangle$ defines an average with respect to
the canonical ensemble with weight $\propto \exp(-{\cal H}/T)$. $A^*$
denotes the complex conjugate of $A$.  For later purposes, the Laplace
transform is defined as follows: 
\begin{align}
\label{laplace.cor}
\Phi(z)&=i\int_0^\infty dt e^{izt}\phi(t) 
= -\langle A^*(\Gamma,0)\frac{1}{z+{\cal L}_+}A(\Gamma,0)\rangle.
\end{align}
The function $\Phi(z)$ is analytic for $\mathrm{Im}(z)>0$.

Following Mori and Zwanzig we can represent the autocorrelation
function in terms of a continued fraction, which is suited to an
approximation by truncation.  
Let us denote the resolvent in Eq.~(\ref{laplace.cor}) by 
${\cal R}=({\cal L}_+ +z)^{-1}$ and
define a scalar product of two dynamic variables $A$
and $B$ via the equilibrium expectation value as
$(A|B):=\langle A^*B\rangle$.
The Laplace transform of the autocorrelation can then be written as a
matrix element of the resolvent
$\Phi(z)= -( A | (z+{\cal L_+})^{-1} A)$.

The idea behind the Mori Zwanzig formalism is to consider explicitly
the dynamic evolution in a subspace of dynamic variables and
treat the dynamics of the remaining variables in a simple
approximation. The crux of the matter is to choose the right
subspace. Popular candidates are the conserved quantities, which are
special because of their long relaxation times in the hydrodynamic
limit. The fast degrees of freedom are then approximated by a single
relaxation time, similar in spirit to the single
collision time approximation of the Boltzmann equation. We shall in
the following consider the subspace spanned by $\rho({\bf r},{\bf
  u})$, but first review the general formalism.

We consider a set of dynamic variables $\{A_i\}$ with normalisation
matrix $(\chi_0)_{ij}=(A_i|A_j)=\langle A_i^*A_j\rangle$.
The projector onto the subspace spanned by the $\{A_i\}$ is given by 
$\hat{P}_0=\sum_{i,j}|A_i) (\chi_0^{-1})_{ij}(A_j|$. The orthogonal
projector is denoted by $\hat{Q}_0=1-\hat{P}_0$.
Following the Mori-Zwanzig formalism we obtain 
an equation for the correlation functions 
$\Phi_{ij}(z)=-(A_i|(z+{\cal L_+})^{-1} A_j)$. In matrix notation it reads
\begin{equation}
\label{phigen.0}
\left(z+ \Omega_0\chi_0^{-1}+ M_0(z)\chi_0^{-1} \right) \Phi(z)
=-\chi_0 
\end{equation}
with $(\Omega_0)_{ij}=(A_i|{\cal L}_+A_j)$ and the memory kernel
\begin{equation}
(M_0)_{ij}(z)=-(A_i| {\cal L}_+ \hat{Q}_0 (z+ \hat{Q}_0{\cal L}_+
    \hat{Q}_0)^{-1} \hat{Q}_0 {\cal L}_+\, A_j).
\end{equation}
Since the memory kernel is again a matrix element of a modified
resolvent taken with the state $B_i=\hat{Q}_0 {\cal L}_+\, A_i$, we can
iterate the procedure, thereby representing the autocorrelation as a
continued fraction. 
We define a second projector 
$\hat{P}_1=\sum_{ij}|B_i)(\chi_1^{-1})_{ij}(B_j|$ and $\hat{Q}_1=1-\hat{P}_1$ 
and apply the same procedure to $M_0(z)$,
\begin{equation}
\label{memgen}
\left(z+ \Omega_1\chi_1^{-1}+  M_1(z)\chi_1^{-1} \right) M_0(z)
=-\chi_1,
\end{equation}
with $(\chi_1)_{ij}=(B_i|B_j), 
(\Omega_1)=(B_i|{\cal L}_+B_j)$ and a second memory kernel $M_1(z)$.

In principle this procedure can be iterated an arbitrary number of
times. However the computation of the static expectation values
$\chi_i$ and $\Omega_i$ becomes increasingly complicated, so that one
ususally truncates the continued fraction after one or two steps. In
the case of hard core interactions the ``restoring forces'' $\Omega_i$
are in general imaginary, because the Liouville operator is non
hermitean. Hence even a truncation of the continued fraction with
$M_1=0$ -- the approximation scheme that will be adopted in this paper -- will give rise to damping effects. This is in contrast to
differentiable potentials with a Hermitean Liouville operator.

\section{Correlations of coupled density orientation fluctuations}
\label{density}
\subsection{Observable and matrix elements}
We are interested here in the anisotropic motion of a tagged rod-like
macromolecule. In particular we want to compute the anisotropic
diffusion constants $D_{\perp}$ and 
$D_{\parallel}$ as well as the
rotational diffusion constant $D_R$ (see Eq.(\ref{DE})) from a kinetic
theory approach. The dynamic variables of interest are thus the
density of a tagged particle $\rho({\bf r},{\bf u})= \delta({\bf
  r}-{\bf r}_0)\delta({\bf u}-{\bf u}_0)$ for all orientations ${\bf
  u}$. It is convenient to consider its Fourier transform $\rho({\bf
  k},{\bf u})= e^{-i{\bf k}{\bf r}_0} \delta({\bf u}-{\bf u}_0)$ and
compute the matrix of correlation functions
\begin{equation}
\Phi_{{\bf u},{\bf u'}}({\bf k},z)=\,-\,(\rho({\bf k},{\bf u})|
(z+{\cal L_+})^{-1}\rho({\bf k},{\bf u'}))
\end{equation}
where the orientation vector ${\bf u}$ is the ``matrix subscript''.
Actually ${\bf u}$ is continuous, so that the matrix equations turn
into integral equations. 
To apply the formalism of Mori and Zwanzig, we identify
\beq
A_{{\bf u}}=
\rho({\bf k},{\bf u})=e^{-i{\bf k}{\bf r}_0}\delta({\bf u}-{\bf u}_0).
\eeq
We consider two steps in the continued fraction and only discuss the
simplest approximations $M_1=0$. As will be shown below, the resulting
correlation function $\Phi_{{\bf u},{\bf u'}}({\bf k},z)$ has the right
hydrodynamic limit. It furthermore accounts for free particle motion
approximately. More elaborate approximations as well as the effects of
futher steps in the continued fraction will be discussed in the
conclusions. 

The computation of susceptibilities $\chi_i$ and ``restoring forces''
$\Omega_i$ is straightforward but cumbersome and hence delegated to
appendix A (see also Refs.~\onlinecite{Hoffman:1969,Verlin:1975}). The results of this calculation are
\begin{align}
(\chi_0)_{{\bf u},{\bf u'}}
&=(A_{\bf u}|A_{\bf u'})
=(4\pi)^{-1} \delta({\bf u}-{\bf u'})\\
(\Omega_0)_{{\bf u},{\bf u'}}&=(A_{\bf u}|{\cal L}_+ A_{\bf u'})=0\\
(\chi_1)_{{\bf u'},{\bf u}}&=
(\dot{A}_{\bf u}|\dot{A}_{\bf u'})
=\frac{1}{4\pi}
\left(
\frac{T}{m}k^2+\frac{T}{I}\hat{L}^2
\right)\delta({\bf u}-{\bf u'})\\
\label{omega1}
(\Omega_1)_{{\bf u'},{\bf u}}
&=
(\dot{A}_{\bf u}|{\cal L}_+ \dot{A}_{\bf u'})
= i \frac{n_0}{8\pi^{5/2}}\left(\frac{T}{m}\right)^{3/2} 
\left[
c_\parallel L^2 ({\bf k}\cdot{\bf u})^2 + 
c_\perp L^2 (k^2-({\bf k}\cdot{\bf u})^2)+
4 \D^2 c_R\hat{L}^2
\right]
\delta({\bf u'}-{\bf u}).
\end{align}
Here the angular momentum operator is defined as usual 
(see Eq.~\eqref{angular_momentum}).
The constants $c_\parallel$, $c_\perp$, and $c_R$ are geometrical factors
specific to needles, depending on the dimensionless ratio $\D=mL^2/(2I)$. They
are defined in App.~\ref{geomfactors} and their numerical values are listed in Tab.~\ref{ctab} for $\D=6$, corresponding to homogeneous rods,
and $\D=2$, corresponding to rods whose mass is concentrated at the endpoints
(``dumbbells'').

\begin{table}
\caption{Geometrical factors $c_\parallel$, $c_\perp$ and $c_R$ for
  homogeneous rods ($\D=6$) and ``dumbbells'' ($\D=2$), rounded to 3 decimal 
  places.}
\label{ctab}
\begin{tabular}{l|l|l|l|l}
\multicolumn{5}{c}{}\\
&\multicolumn{2}{c|}{smooth rods}&\multicolumn{2}{c}{rough rods}\\
 &$\D=6$&$\D=2$&$\D=6$&$\D=2$ \\ \hline
$c_\parallel$ & 0 & 0 & 10.631 & 10.202 \\
$c_\perp$ & 7.253 & 8.626 & 11.802 & 13.390 \\
$c_R$ & 0.544 & 0.684 & 0.871 & 1.058
\end{tabular}
\end{table}

\subsection{The correlation function}
Given the matrix elements obtained above, we proceed to find a
solution for $\Phi_{{\bf u},{\bf u'}}({\bf k},z)$ within the
approximation $M_1=0$.  Since $\Omega_0$ and $\chi_0$ are trivial
quantities, we will concentrate on $\Omega_1$ and $\chi_1$ and drop
the subscript 1 from now on for notational simplicity. Additionally,
we absorb a factor of $1/\chi_0=4\pi$ by defining $\varphi=4\pi\Phi$.
Multiplying Eq.~(\ref{phigen.0}) from the left by $(z+\Omega\chi^{-1})$
(and using Eq.~(\ref{memgen})), we are then left with the matrix
equation (w.r.t.\ ${\bf u}$ and ${\bf u'}$)
\begin{equation}
\label{matrixeq}
(z^2+z \Omega\chi^{-1}-4\pi \chi)\varphi=-(z+\Omega\chi^{-1})
\end{equation}
What is an appropriate set of functions to represent $\varphi$ in? The matrix
$\chi$ is diagonal in the eigenfunctions of the isotropic Laplacian, 
i.e.\ the spherical harmonics.  The matrix $\Omega$ is diagonal in the
eigenfunctions of the anisotropic Laplacian, denoted by $y_{lm}(ic, {\bf u})$
and related to the {\it oblate} spheroidal polynomials (or {\it oblate}
spheroidal wave functions) $S_{lm}(ic,\eta)$ in the same way as the spherical
harmonics are related to the associated Legendre functions,
\begin{align}
\label{spheroidal1}
y_{lm}(ic, {\bf u}) &= y_{lm}(ic, \eta, \phi)
=\frac{1}{\sqrt{2\pi}}S_{lm}(ic,\eta)e^{im\phi},
\end{align}
with $\eta={\bf u}\cdot {\bf k}/k$. The properties of 
spheroidal wave functions are discussed in App.~\ref{spheroidal}.

As will become evident below, we do not need to keep the dependence on
the azimuthal angles $\phi$ and $ \phi'$ in Eq.~\eqref{matrixeq} 
and integrate the equation
over these variables to obtain
\begin{align}
\label{matrix2eq}
\int d\eta''\,
(z^2+z \Omega\chi^{-1}-4\pi \chi)_{\eta\eta''}\varphi_{\eta''\eta'}
&=-(z+\Omega\chi^{-1})_{\eta\eta'}.
\end{align}
Here $\varphi_{\eta\eta'}$ is defined as
\beq
\varphi_{\eta\eta'}=\frac{1}{2\pi}\int d\phi\int d\phi'\,
\varphi_{{\bf u}{\bf u}'}
\eeq
and analogously
\begin{align}
\begin{split}
\Omega_{\eta\eta'}&=
  \sum_l \Omega_l S_{l0}(ic,\eta) S_{l0}(ic,\eta')\\
\chi_{\eta\eta'}&=  \sum_l \chi_l S_{l0}(0,\eta) S_{l0}(0,\eta').
\label{OmegaChi}
\end{split}
\end{align}
The expansion coefficients are given by
\begin{align}
\begin{split}
\Omega_l &= i\kappa\left(
  \frac{c_{\perp}k^2 L^2}{4 \D^2 c_R}+\lambda_{l0}(-c^2)\right),\\
\chi_l &= \frac{1}{4\pi}\left(\frac{T}{m}k^2
+\frac{T}{I}l(l+1)\right),
\label{OmegaChiDef}
\end{split}
\end{align}
with $c^2=\frac{c_{\perp}-c_{\parallel}}{4\D^2c_R}(L k)^2$ and $\kappa
=\frac{\rho_0 \D^2 c_R}{2\pi  L^3}\left(\frac{T}{\pi m}\right)^{3/2}$.

In App.~\ref{diag} we show how to diagonalize
$\varphi_{\eta\eta'}$ perturbatively for $k\to 0$.
For the time being, however, we choose the set
$S_{l0}(0,\eta) \propto P_l(\eta)$ to expand $\varphi_{\eta\eta'}$ as
\beq
\label{exp.legendre}
\varphi_{\eta\eta'} =\sum_{l,l'} \varphi_{ll'}
S_{l0}(0,\eta) S_{l'0}(0,\eta').
\eeq
In particular, we will later need the matrix elements $\varphi_{00}$
and $\varphi_{20}$, which are calculated from the diagonalized $\varphi$ in
App.~\ref{diag}. The result is
\beqa
\label{phi00}
\varphi_{00}&=&
-\frac{z+\chi_0^{-1}\Omega_0}{z^2+z\chi_0^{-1}\Omega_0-4\pi\chi_0}\\
\label{phi02}
\varphi_{20}&=&4\pi\frac{c^2}{9\sqrt{5}}\frac{-\Omega_2}
{\left(z^2+z\chi_0^{-1}\Omega_0-4\pi\chi_0\right)
\left(z^2+z\chi_2^{-1}\Omega_2-4\pi\chi_2\right)}\nonumber\\
&=&\varphi_{02} + \mathcal{O}(k^4),
\eeqa
where only terms up to and including ${\cal O} (k^2)$ have been retained.

\subsection{Numerical simulations}
\label{sec:num}
We have performed simulations of the system of hard rods using an
event-driven algorithm as described in Ref.~\onlinecite{huthmann:99} (with no
inelasticity). We used system sizes of $N=800$ to $N=2048$ and
densities between $\rho_0=0.25$ and $\rho_0=100$.  For the most part we
used homogeneous rods with $\D=6$. The results of the simulations will be discussed below along with the theoretical ones.

\section{Results}
\label{sec:results}
In this section we are going to discuss the results of the approximate
kinetic theory obtained by setting $M_1=0$. One might expect that our theory
reduces to the diffusion equation of Perrin in the limit of
small wave numbers $kL\ll 1$ and long times. However, such 
a straightforward limit does not exist as we will show below.

There are two special cases where the comparison between microscopic theory
and the Perrin equation is straightforward.  One case concerns
purely rotational motion. We show that the theory predicts a crossover as a
function of density from weakly damped free rotations to rotational diffusion.
We furthermore obtain explicit expressions for the rotational diffusion
constant of the Perrin theory as well as for the damping of the free
rotations.  The other important case which allows for direct comparison with
the Perrin theory is purely translational motion, where we recover the
isotropic diffusion constant.

The third and most interesting case, however, concerns the coupling of
translational and rotational degrees of freedom. The comparison with
the Perrin diffusion equation is not directly evident. In fact, we
derive -- within kinetic theory -- the full time dependence of the
mean square displacements of a tagged rod parallel and perpendicular
to its initial orientation. We then compare our expressions to the
corresponding ones obtained from the Perrin theory. An important
problem arising in the analysis is the appearance of short time scales
present in the kinetic theory which are not easily decoupled from
the long-time behavior. One would like to map the latter to the Perrin
diffusion equation which involves only long-time diffusive time
scales. Even within the Perrin theory, the anisotropic translational
motion characterized by the diffusion constants $D_\parallel$ and
$D_\perp$ is only perceptible on small times (compared to the
rotational diffusion time) before the anisotropy is hidden by the
long-time isotropic translational diffusion of the center of mass of
the tagged rod.  On the side of kinetic theory, the time interval for
anisotropic diffusion to be observed is bounded from below by
ballistic motion which is isotropic a priori.  We comment on this
point in more detail below when comparing our analytical results with
simulations.

\subsection{Rotational diffusion}
\label{rotdif}

We first discuss purely rotational motion, as obtained from the
results of Sec.~\ref{density} by setting ${\bf k}=0$. In order to
compare with simulation results later on we focus on the Fourier
transform $\varphi(\omega)$ instead of the Laplace transform
$\varphi(z)$ of the correlation function. Since in the time domain
$\varphi(t)$ is a real symmetric function, the Fourier and Laplace
transforms are related via
$\varphi(\omega) = 2\mathrm{Im}(\varphi(z=\omega+i0))$
where $\omega$ is real.

The correlation function for $\mathbf{k}=0$ is diagonal in the angular
momentum quantum numbers and is given by
\begin{align}
%\varphi_{l}^{''}(\omega)=\frac{\zeta \omega_l^2}
%{(\omega^2-\omega_l^2)^2+\zeta^2\omega^2}
\varphi_{ll}(z) &= 
-\frac{z^*|z+i\zeta|^2-(z+i\zeta)\omega_l^2}
{|z^2-\omega_l^2|^2+|z|^2\zeta^2+2\mathrm{Im}(\zeta z^*(z^2-\omega_l^2))}
\end{align}
with
$\omega_l^2=\frac{T}{I}l(l+1)$ and $\zeta=4\pi\kappa\frac{I}{T}$.
The Fourier transform is
\begin{align}
\label{phiq=0}
  \varphi_{ll}(\omega) &= 
  2\frac{\zeta\omega_l^2}{(\omega^2-\omega_l^2)^2+\omega^2\zeta^2}.
\end{align}

The correlation function Eq.~(\ref{phiq=0}) has two poles in the lower
complex frequency plane at $\omega=-i\zeta/2\pm
\sqrt{(\omega_l^2-\zeta^2/4)}$ and the corresponding complex
conjugates in the upper half plane. It shows qualitatively different
behaviour depending on the reduced density $\rho_0=n_0 L^3$. For low
density the poles have a nonzero real part, corresponding to weakly
damped oscillatory motion of the rods:
\begin{equation} 
\varphi_{ll}(t)=\frac{1}{\omega_l}\exp{(-\zeta t/2)} 
\left(\omega_l\cos{(\omega_l t)}+\zeta\sin{(\omega_l t)}\right) 
\end{equation} 

The real part of the poles vanishes above a critical density 
$\rho_{\text{crit}}(l)=\left( l(l+1)\frac{8 \pi^3 }{c_R^2 \D} \right)^{1/2}$
which depends on $l$. For large density the time dependent correlation 
function is a sum of two exponentials with the long relaxation time 
$\tau_{\text{long}}(l)=(D_R l(l+1))^{-1}$ and the short relaxation time 
$\tau_{\text{short}}=\zeta^{-1}$. The decay on long times corresponds to
rotational diffusion and is in agreement with the Perrin theory. 
The rotational diffusion constant is given by 
\begin{equation} 
\label{DRdef}
D_R= \frac{2 \pi^{3/2}}{\rho_0 L c_R}\sqrt{\frac Tm} 
\end{equation} 

Our approximate theory does not describe the free particle limit 
$\rho_0 \to 0$ correctly. In particular the weakly damped oscillations
acquire additional damping due to free particle motion (sometimes 
called motional narrowing). We discuss a simple way to account for 
these additional terms approximately. 

If there are no interactions, the angular correlations are simply
given by 
\begin{equation} 
\varphi_{ll}^0(t)=\langle P_l(\cos{\omega t})\rangle_{\text{th}}=\int_0^\infty d\omega\; 
\omega\frac{I}{T}\; 
P_l(\cos{\omega t})\; e^{-\frac{\omega^2 I}{2 T}}.
\end{equation} 
For example for $l=1$ (the case we will need below) 
the correlation function $\varphi_{11}^0(z)$ can be expressed in terms
of the exponential integral function $E_1(z)$ \cite{abramowitz}.

A simple way to incorporate free particle motion is the following. One
represents the correlation function $\varphi_{ll}^0$ as a continued 
fraction with $\Omega=0$ but $M\neq 0$ (see Eq.~\eqref{phigen.0}),
\begin{equation}
\varphi_{ll}^0(z)=-\frac{1}{z-\frac{4\pi\chi_l}{z+ M_l/\chi_l}}.
\end{equation}
Then the memory kernel is expressed in terms of $\varphi_{ll}^0$,
\begin{equation}
z+M_l /\chi_l=4\pi\chi_l\frac{\varphi_{ll}^0(z)}{z\varphi_{ll}^0(z)+1},
\end{equation}
and can now substituted in the continued fraction for the interacting system.
This procedure yields 
\begin{equation} 
\varphi_{ll}(z)=
-\frac{\frac{\Omega_l}{\chi_l}(z\varphi_{ll}^0 +1)+4\pi\varphi_{ll}^0 \chi_l}
{z\frac{\Omega_l}{\chi_l}(z\varphi_{ll}^0 +1)+4\pi\chi_l}.
\label{phiuueq}
\end{equation}

We now compare the Fourier transform $\varphi_{ll}(\omega)=
2\mathrm{Im}(\varphi_{ll}(z=\omega+i0))$ for the case $l=1$, which
corresponds to the correlation function $\varphi_{11}(t)=\langle{\bf
  u}_0(t)\cdot {\bf u}_0(0)\rangle$, with numerical simulations in
Fig~\ref{phiuufig}. 
Agreement is very good for small
densities because the free particle limit has been incorporated. Our
approximate theory also reproduces the crossover from damped
oscillations, corresponding to a peak at finite frequency in the
spectrum, to purely relaxational motion, characterized by a peak at
zero frequency.

\subsection{Translational dynamics}
\label{transdif}
Next we discuss purely translational motion, obtained from the general
results by setting $l=l'=0$, i.e.\ by examining the correlation function
$\varphi_{00}$ as given in Eq.~\eqref{phi00}.
Its Fourier transform is given by 
\begin{align}
\varphi_{00}(\omega)&=
2\frac{(k v_{\text{th}})^2}{(\omega^2-(k v_{\text{th}})^2)^2+\xi^2\omega^2}
\end{align}
with $\xi=\nu \frac{c_{\parallel}+2c_{\perp}}{(2\pi)^2}$.
Here we have introduced the Enskog collision frequency
$\nu=\frac{2\rho_0}{3L}\sqrt{\frac{\pi T}{m}}$
and the thermal velocity $v_{\text{th}}^2=T/m$.

The corresponding time dependent correlation function is the sum of
two exponentials.  The decay rate of the fast component,
$\tau_{0}$, is determined by the collision rate according to
$\tau_{0}^{-1}=\nu(c_{\parallel}+2c_{\perp})/(2\pi)^2$. The decay
rate of the slow component is diffusive as it should be, due to
particle number conservation. Since we consider translational motion
only and consequently have set $l=0$ in the equation of motion, we
should recover the isotropic diffusion constant of the Perrin theory.
The latter predicts $D_{\text{iso}}=(D_{\parallel}+2 D_{\perp})/3$.
Comparison with the Perrin theory allows us to identify 
\begin{align}
\label{Disodef}
D_{\text{iso}}&=\frac{v_{\text{th}}^2}{\nu}\frac{(2\pi)^2}{c_{\parallel}+
2c_{\perp}} = \sqrt{\frac{T\pi}{m}}
\frac{6\pi L}{\rho_0 (c_\parallel+2c_\perp)}.
\end{align}

The ratio of the rotational to the isotropic translational diffusion constant
is given in the Kirkwood-Riseman theory \cite{riseman} by
$D_R/D_{\text{iso}}=9/L^2$. Even though this theory is designed for a rodlike
molecule in a solvent where hydrodynamic interactions are important, we do
find almost the same result for a melt of smooth needles with a homogeneous
mass distribution: $D_R/D_{\text{iso}}=(c_{\parallel}+2c_{\perp})/(3c_R
L^2)=8.89/L^2$ (see Tab.~\ref{ctab}). The ratio does depend on the mass
distribution though and is different for smooth and rough rods.

In summary, our approximate kinetic theory correctly describes
particle diffusion which dominates the long time decay of the
translational correlation function. It yields an approximate
expression for the isotropic diffusion constant, which is inversely
proportional to the density $\rho_0$.
The decay of the nonconserved variables
has been approximated by a single relaxation time, which is inversely
proportional to the Enskog collision frequency.

\subsection{Coupled translational and rotational dynamics: kinetic theory}
\subsubsection{From correlation functions to mean square displacements}

Our next goal is to calculate the mean-square displacement of the tagged rod,
decomposed into the directions parallel and perpendicular to its initial
orientation $\mathbf{u}_0(0)$. We define the tensor
\begin{align}
  R_{\alpha\beta\gamma\delta} &= \langle\Delta r_{\alpha}(t)\Delta
  r_{\beta}(t) u_{0\gamma}(0) u_{0\delta}(0) \rangle,
\end{align}
where $\Delta\mathbf{r}(t)=\mathbf{r}_0(t)-\mathbf{r}_0(0)$ is the
displacement of the tagged rod.
% the total and directional mean-square
%displacements can be written as
%\begin{align}
%\label{msd1}
%\langle \Delta\mathbf{r}^2(t)\rangle &=
%\sum_{\alpha\beta}R_{\alpha\alpha\beta\beta} \\
%\label{msd2}
%\langle \Delta r_{\parallel}^2(t)\rangle &=
%\sum_{\alpha\beta}R_{\alpha\beta\alpha\beta} \\
%\label{msd3}
%\langle \Delta\mathbf{r}_{\perp}^2(t)\rangle &=
%\langle \Delta\mathbf{r}^2(t)\rangle - 
%\langle \Delta r_{\parallel}^2(t)\rangle.
%\end{align}
This tensor $R_{\alpha\beta\gamma\delta}$ is isotropic and
symmetric under exchange of the indices $\alpha$ and $\beta$ and also of
$\gamma$ and $\delta$. These symmetries imply that there are only two
independent tensor components $R_1$ and $R_2$ such that
\begin{align}
\label{tensorsym}
R_{\alpha\beta\gamma\delta} &= \delta_{\alpha\beta}\delta_{\gamma\delta}
R_1 +(\delta_{\alpha\gamma}\delta_{\beta\delta}+
\delta_{\alpha\delta}\delta_{\beta\gamma}) R_2.
\end{align}
In order to calculate $R_{\alpha\beta\gamma\delta}$, we express it in terms
of the correlation function $\Phi$,
\begin{align}
\begin{split}
R_{\alpha\beta\gamma\delta} 
&= -\frac{\partial^2}{\partial k_\alpha \partial k_\beta}
\left\langle 
e^{i\mathbf{k}\cdot\Delta\mathbf{r}(t)}u_{0\gamma}(0) u_{0\delta}(0)
\right\rangle\Big|_{\mathbf{k}=0}\\
&=-\frac{\partial^2}{\partial k_\alpha \partial k_\beta}
\int d\mathbf{u}\,d\mathbf{u}'\,u'_\gamma u'_\delta 
\Phi_{\mathbf{u}\mathbf{u}'}(\mathbf{k},t)
\Big|_{\mathbf{k}=0}.
\end{split}
\end{align}
The total mean-square displacement can easily be expressed in terms of
this tensor as
\begin{align}
\label{Raabb}
\langle \Delta\mathbf{r}^2(t)\rangle = \sum_{\alpha\beta}R_{\alpha\alpha\beta\beta}
= 9R_1 + 6R_2 = 
-\sum_\alpha \frac{\partial^2}{\partial k_\alpha^2} 
\int d\mathbf{u}\,d\mathbf{u}' \Phi_{\mathbf{u}\mathbf{u}'}(\mathbf{k},t)
\Big|_{\mathbf{k}=0} =
 -3 \frac{\partial^2}{\partial k^2}\varphi_{00}\Big|_{k=0}
\end{align}
(recall that $\Phi$ and $\varphi$ are identical up to a factor of $4\pi$). In
the last step the derivatives become simpler since $\varphi_{00}$ is only a
function of $k^2$ (cf.\ Eq.~\eqref{phi00}).

In terms of $R_1$ and $R_2$ the parallel mean-square displacement
is given by (cf.\ Eq.~\eqref{tensorsym})
\begin{align}
\label{rpar}
\langle \Delta r_{\parallel}^2(t)\rangle &= \sum_{\alpha\beta}
R_{\alpha\beta\alpha\beta} = 3R_1 + 12R_2.
\end{align}
In order to evaluate this further, we need to know one more equation besides
Eq.~\eqref{Raabb} to determine $R_1$ and $R_2$. This is provided by
\begin{align}
\begin{split}
\sum_\alpha R_{\alpha\alpha\alpha\alpha} &= 3R_1 + 6R_2 =
-\sum_\alpha \frac{\partial^2}{\partial k_\alpha^2}
\int d\mathbf{u}\,d\mathbf{u}' (u'_\alpha)^2 \Phi_{\mathbf{u}\mathbf{u}'}
(\mathbf{k}, t)\Big|_{\mathbf{k}=0} \\
&= -\frac{\partial^2}{\partial k^2}
(\frac{2}{\sqrt{5}}\varphi_{02}+\varphi_{00})\Big|_{k=0}.
\label{Raaaa}
\end{split}
\end{align}
Note that each term in the sum over $\alpha$ does not depend on $\alpha$
because of symmetry and the fact that $\varphi_{00}$ and $\varphi_{02}$ only
depend on $k^2$. Thus it suffices to evaluate the term with e.g.\ $\alpha=3$.

Eqs.~\eqref{Raabb}, \eqref{rpar}, and \eqref{Raaaa} then result
in
\begin{align}
\label{rpareqn}
\langle \Delta r_{\parallel}^2(t)\rangle &= 
-\frac{\partial^2}{\partial k^2}
(\varphi_{00}+\sqrt{5}\varphi_{02})\Big|_{k=0} \\
\label{rperpeqn}
\langle \Delta\mathbf{r}_{\perp}^2(t)\rangle &=
-\frac{\partial^2}{\partial k^2}
(2\varphi_{00}-\sqrt{5}\varphi_{02})\Big|_{k=0}.
\end{align}

\subsubsection{Anisotropic diffusion from kinetic theory} 

Given the expression derived above for the mean square displacement parallel
and perpendicular to the initial orientation, the last step remaining is the
calculation of the correlation functions $\varphi_{00}(k,t)$ and
$\varphi_{20}(k,t)$ in the time domain from the corresponding expressions
derived in the previous section for the Laplace transform (depending on $z$).
Therefore, we need to analyse the poles of $\varphi_{00}(k,z)$ and
$\varphi_{20}(k,z)$. The denominators of $\varphi_{00}(k,z)$ and
$\varphi_{20}(k,z)$ consist of factors of the form
$z^2+\frac{\Omega_n}{\chi_n} z-4\pi \chi_n$  ($n=0,2$).  The two roots of
each of these expressions are at
\begin{align}
z_{n,\pm} &= -\frac{\Omega_n}{2\chi_n}(1\pm\sqrt{1+16\pi\chi_n^3/\Omega_n^2}).
\end{align}
While $\varphi_{00}(k,z)$ only contains the $n=0$ factor and thus has the two
poles $z_{0,\pm}$, $\varphi_{20}(k,z)$ has both the $n=0$ and $n=2$ factors
and therefore has all four poles.

Given these poles, we can expand $\varphi_{00}$ and $\varphi_{20}$ from
Eqs.~\eqref{phi00} and \eqref{phi02} into partial fractions and carry out the
inverse Laplace transforms. From Eqs.~\eqref{rpareqn} and \eqref{rperpeqn} we
then get the mean-square displacements. Before we show the results of this
straighforward but tedious calculation, we note that the long-time behavior on
long length scales ($\mathbf{k}\to 0$) is determined by the pole or poles
closest to the origin. 
Each pole $z_{n,\pm}$ corresponds to a timescale $1/(iz_{n,\pm})$ (provided
$z_{n,\pm}$ is purely imaginary, see below). Inserting the values from
Eq.~\eqref{OmegaChiDef} these timescales are (immediately setting $k=0$ 
where possible)
\begin{align}
\tau_{0} &= \frac{1}{iz_{0,+}} = D_{\text{iso}}\frac{m}{T} =
  \sqrt{\frac mT} \frac{6L\pi^{3/2}}{\rho_0(2c_\perp+c_\parallel)}\\
\tau_{1} &= \frac{1}{iz_{2,+}} =
  \sqrt{\frac mT}\frac{2\pi^{3/2}L}{\rho_0 \D c_R}
  \left(1+\sqrt{1-48\pi^3/(\rho_0^2 \D c_R^2)}\right)^{-1}\\
\tau_{2} &= \frac{1}{iz_{2,-}} = 
  \sqrt{\frac mT}\frac{2\pi^{3/2}L}{\rho_0 \D c_R}
  \left(1-\sqrt{1-48\pi^3/(\rho_0^2 \D c_R^2)}\right)^{-1}\\
\tau_{3} &= \frac{1}{iz_{0,-}} = \frac{1}{D_{\text{iso}}k^2}.
\end{align}
Generically, only one of these times is long, namely $\tau_3$,
which is proportional to $1/k^2$.  For
large densities, however, $\tau_2$ becomes proportional to the density and
thus large, too.  This indicates
that, at least for high densities, there are \textit{two} long time scales.
The other two times scales, $\tau_{0}$ and $\tau_{1}$, are short.
The first one, $\tau_{0}$, is related to the
Enskog collision time.
The second one, $\tau_1$, is of the same order as $\tau_0$.
For large densities it agrees with the short rotational timescale
$\tau_{\text{short}}$ from Sec.~\ref{rotdif}. It characterizes the
crossover from free rotational motion to rotational diffusion.

After performing the inverse Laplace transforms, we find
\begin{align}
\langle \Delta\mathbf{r}^2\rangle &= 6D_{\text{iso}}
\left(t+\tau_0(\exp(-t/\tau_0)-1)\right) \\
\langle \Delta r_\parallel^2\rangle &= 
  \frac 13 \langle \Delta \mathbf{r}^2\rangle + 
\langle \Delta r_{\text{aniso}}^2 \rangle \\
\langle \Delta \mathbf{r}_\perp^2\rangle &= 
  \frac 23 \langle \Delta \mathbf{r}^2\rangle -
\langle \Delta r_{\text{aniso}}^2 \rangle
\end{align}
where
\begin{align}
\langle \Delta r_{\text{aniso}}^2\rangle &= 
-\frac{4\pi\kappa (c_\perp-c_\parallel)L^2}{3\D^2 c_R}
\sum_{p=0}^3\left.
\frac{e^{-i t/\tau_p}}{\prod_{p'\ne p}(1/\tau_p-1/\tau_{p'})}
\right|_{k=0}.
\label{raniso}
\end{align}
Due to the density dependence of $\Omega_2$ the poles $z_{2,\pm}$
acquire a real part for \textit{small} densities, such that the pole
is no longer purely diffusive but oscillatory.  Note, however, that
the mean-square displacements are real quantities whether
$z_{2,\pm}$ are purely imaginary (corresponding to an exponential
decay of the nonlinear part of the mean-square displacements) or not
(corresponding to an oscillatory behavior). 

The anisotropic mean square displacements calculated from
kinetic theory have the following important characteristics:
\begin{itemize}
\item
For short times $t\ll \tau_0$ one obtains ballistic behavior
$2\langle\Delta r_{\parallel}^2(t) \rangle\simeq
\langle\Delta \mathbf{r}_{\perp}^2(t)\rangle\simeq
2v_{\text{th}}^2t^2$
The time $\tau_0$ is identical to the fast timescale calculated in 
Sec.~\ref{transdif}.
\item For long times $t\gg \tau_2$ the mean square diplacements
  approach the isotropic center of mass diffusion behavior
$2\langle\Delta r_{\parallel}^2(t) \rangle\simeq
\langle\Delta \mathbf{r}_{\perp}^2(t) \rangle\simeq
4D_{\text{iso}}t$
\item For large densities, the pole $z_{2,-}$ is given by
\begin{align}
z_{2,-} &\simeq -\frac{\Omega_2}{2\chi_2}\left(
1-1-8\pi\frac{\chi_2^3}{\Omega_2^2}
\right)
=-6i D_R
\end{align}
with the rotational diffusion constant $D_R$ given in Eq.~\eqref{DRdef}.
The timescale $\tau_2$ is therefore $\tau_2 = (6D_R)^{-1} =
\tau_{\text{long}}(l=2)$ 
where $\tau_{\text{long}}$ is the long rotational timescale (see 
Sec.~\ref{rotdif}). It also agrees with the crossover time from Perrin theory,
see Eqs.~\eqref{displacementpar} and \eqref{displacementperp}.
\item
For intermediate times, when $ \tau_0 \ll t \ll
\tau_2$ (provided such an interval exists, which is only the case for
high enough density), there is a linear (in $t$) regime in $\langle \Delta
r_{\text{aniso}}^2\rangle$ which corresponds to \textit{anisotropic
  diffusion}. The diffusion constants $D_\perp=D_{\text{iso}}-\frac 12 \Delta D$ and
$D_\parallel=D_{\text{iso}}+\Delta D$ in this regime can be calculated from
Eq.~\eqref{raniso}, resulting in
\begin{align}
\begin{split}
  \Delta D &= \frac{2\pi\kappa(c_\perp-c_\parallel)L^2}
{3\D^2c_R(1/\tau_2-1/\tau_1)(1/\tau_2-1/\tau_0)}\\
&\stackrel{\rho_0\to\infty}{\longrightarrow}
\frac{(c_\perp-c_\parallel)L^2}{(2c_\perp+c_\parallel) \D}D_R
\label{deltad}
\end{split}
\end{align}
\end{itemize}

\subsubsection{Anisotropic diffusion from simulations}

The results of our analytic calculation are compared with the simulations
in Figs.~\ref{r2figD6a}, \ref{r2figD6b} and \ref{r2figD6c} for homogeneous rods ($\D=6$) and in Fig.~\ref{r2figD2} for dumbbell molecules ($\D=2$). 

The agreement between theory and simulations is very good for small densities
$\rho_0\le 10$ but at these densities there is hardly any discernible
anisotropy (see Fig.~\ref{r2figD6a}).

For higher densities an anisotropy window opens up
between the parallel and perpendicular mean-square displacement curves. This
is exemplified in Fig.~\ref{r2figD6b}. For densities above $\approx 20$ the
isotropic diffusion constants (slopes of the mean square displacements)
\textit{increase} with density in the simulation. This effect is well known
\cite{frenkel:83,magda:86} and is due to enhanced parallel diffusion along a
tube formed by surrounding particles. This involves correlated particle
collisions and is thus not captured within our theory. Therefore the theory does not fit the data quantitatively but Fig.~\ref{r2figD6b} shows that the anisotropy is very well captured qualitatively. But even though anisotropy is clearly present, it can (at these densities) not adequately be described by anisotropic diffusion constants since the anisotropy window is still small and within it, no convincing ``straight line fit'' appears possible. This is true for both theory and simulation.

Fig.~\ref{r2figD6c} shows a direct comparison of theory and simulations of the mean-square displacements for densities $\rho_0=40$ and $\rho_0=100$. Here it becomes obvious that the isotropic diffusion constant is underestimated by the theory. The simulations at $\rho_0=100$ now very clearly show anisotropic diffusion with a well-defined perpendicular diffusion constant $D_\perp$. The parallel diffusion constant $D_\parallel$, on the other hand, is still not well defined even at this density.

The anisotropy becomes even more pronounced for dumbbell molecules with $\D=2$
at $\rho=100$, which is shown in Fig.~\ref{r2figD2}. The reason the
effect is more pronounced for dumbbells lies in the $\D$-dependence of
the diffusion constants. While $\Delta D$ is proportional to $1/\D$
(see Eq.~\eqref{deltad}), $D_{\text{iso}}$ is only weakly dependent on $\D$
through $c_\parallel$ and $c_\perp$. Therefore
anisotropic diffusion is more pronounced for smaller $\D$.

\section{Discussion and Outlook}
\label{sec:diss}
Before we go into a discussion of the details of our results we will give a brief overview of our main findings.
The aim of this paper is to gain some microscopic understanding of the transport properties of a gas of thin hard rods beyond the phenomenological Perrin equation. Purely rotational and translational dynamics of a tagged particle can be described by simple rotational or translational diffusion, in agreement with the Perrin equation, allowing us to calculate the rotational and isotropic translational diffusion constants. The motion of a tagged particle along and perpendicular to its original orientation shows marked anisotropies. The nature of the anisotropy is, however, very different from the one predicted by Perrin theory, at least for low and medium densities. This is confirmed by simulations.  The Perrin equation predicts anisotropies which can be described as anisotropic diffusion for all densities. In contrast, our results for low densities exhibit only a very drawn-out crossover from ballistic motion for short times to isotropic diffusion at long times with no intermediate regime which could be characterized as anisotropic diffusion. For medium densities $\rho_0\approx 40$ such a regime begins to emerge; it is however still far from being well-defined. Even for densities $\rho_0= 100$ where the mean-square displacement is clearly anisotropic, anisotropic \textit{diffusion} is only seen in the perpendicular motion. 

We have derived an approximate theory for the coupled
translational and rotational motion of thin hard rods, starting from a
microscopic model. The dynamic
evolution is determined by binary collisions, in which the normal
component of the velocity of the contact point is reversed and the
tangential component is either left unchanged (smooth rods) or
reversed as well (rough rods). The Mori-Zwanzig projection operator
formalism has been applied to the correlation of a tagged rod's
density $\rho({\bf r},{\bf u},t)$, specifying its position ${\bf r}$
and orientation ${\bf u}$ at a given time $t$. Matrix elements of the
collision operator have been computed, resulting in a generalized
Enskog theory, in which only uncorrelated binary collisions are taken
into account -- similar in spirit to a ``Sto\ss zahl Ansatz'' of the
corresponding Boltzmann equation. Two steps in a continued fraction
expansion give rise to an integral equation for the correlation
function. This equation has been solved in the limit of small
wavenumber, where it is rendered diagonal by linear combinations of
the spherical harmonics. We have also performed numerical simulations,
using an event driven algorithm for hard needles.

Purely rotational motion (${\bf k}=0$) is as expected and in agreement with
previous discussions. For small density we observe damped oscillations with a
frequency $\omega^2=T/I$. In the dilute limit it is essential to include free
particle motion, which provides the dominant damping mechanism.

For large densities the dynamics is purely relaxational. The crossover
happens at a critical density $\rho_{\text{crit}}^2(l)=8\pi^3 l(l+1)/(c_R \D)$
which depends on the angular momentum $l$, the moment of inertia via
$\D=\frac{mL^2}{2I}$ and on whether we are considering smooth or rough
rods. For $\rho_0\geq \rho_{\text{crit}}(l)$ our approximate kinetic theory
predicts two relaxation times. The longer one is determined by the
rotational diffusion constant according to $\tau_{\text{long}}(l)=(D_R
l(l+1))^{-1}$ with the rotational diffusion constant $D_R$ given by
Eq.~\eqref{DRdef}.
It depends on the mass distribution along the rod and
is different for smooth and rough needles. The rotational diffusion
constant $D_R$ is always larger for smooth rods than for rough rods,
e.g. for a homogeneous mass distribution along the rod we find
$D_R^{\text{smooth}}/D_R^{\text{rough}}\approx 1.60$.

The long time relaxation $\tau_{\text{long}}(l)=(D_R l(l+1))^{-1}$ is in
agreement with the Perrin equation, where $D_R$ enters as a
phenomenological parameter. In addition our theory also predicts a
fast decay with a relaxation rate $\tau_{\text{short}}=D_R I/T$ or
alternatively $\frac{\tau_{\text{short}}}{\tau_{\text{long}}}=
\frac{D_R^2I}{T}l(l+1)=(\frac{\rho_{\text{crit}}}{2\rho_0})^2$.  This short
relaxation time is mainly determined by the moment of inertia and
hence roughly three times larger for dumbbells than for a homogeneous
mass distribution.

Next we discussed purely translational motion ($l=0$).
We find relaxation with a short and a long relaxation time, the
latter being diffusive $\tau_l=(D_{\text{iso}}k^2)^{-1}$ and in agreement
with the Perrin equation, which allowed us to identify the isotropic
diffusion constant $D_{\text{iso}}$ as in Eq.~\eqref{Disodef}.
It depends on the mass distribution and is different for smooth and rough
needles.  The
isotropic diffusion constant is always larger for smooth needles than
for rough ones, e.g. for a homogeneous mass distribution along the rod
we find
$D_{\text{iso}}^{\text{smooth}}/D_{\text{iso}}^{\text{rough}}\approx 2.36$.
Comparing different mass distributions we always find a larger
diffusion constant for the homogeneous mass distribution than for the
dumbbells. In addition to the diffusive long time decay, the kinetic
theory predicts partial decay on a microscopic timescale, which is
given by $\tau_0=D_{\text{iso}}m/T$.  

The most interesting case is of course the coupled translational and
rotational motion, as described by the time dependent correlation
$\Phi_{{\bf u},{\bf u'}}({\bf k},t)$.
This function has been diagonalised in the long wavelength limit by
linear combinations of spherical harmonics, which for general ${\bf
  k},l,m$ differ from the solution to the Perrin equation. We
attribute this difference to the fact that the Perrin equation is {\it not}
simply a gradient expansion but involves the angular momentum
operator, which does not contain any small parameter allowing for a
systematic expansion. 

Anisotropic diffusion is best discussed with the help of the particle's
displacement vector, projected onto or perpendicular to the particle's
initial orientation, i.e.\ $\langle \Delta r_{\parallel}^2(t)\rangle $
and $\langle \Delta r_{\perp}^2(t)\rangle$. In general,
we find three dynamic regimes. For short times $t\ll \tau_0$ the
motion is ballistic. For the longest timescales, i.e. times larger
than the timescale for rotation of a needle, $t\gg 1/(6D_R)$,
isotropic diffusion prevails. In between a time regime appears which
is characterized by anisotropic diffusion: a linear increase of the
mean square displacements $\langle \Delta r_{\parallel}^2(t)\rangle
\sim D_{\parallel}t$ and $\langle \Delta r_{\perp}^2(t)\rangle \sim
D_{\perp}t$ with anisotropic diffusion constants $D_{\parallel}$ and
$D_{\perp}$. To clearly see this intermediate regime, the density has
to be sufficiently large, such that the timescale for rotational
diffusion of the rods is long. In the simulations these three time
regimes are clearly visble for densities $\rho_0\geq 40$ (see
Fig.~\ref{r2figD6b}).  The analytical theory correctly predicts
anisotropic diffusion qualitatively (see Fig.~\ref{r2figD6b}), however
for such high densities no quantitative agreement between simulations
and theory is achieved, see Fig.~\ref{r2figD6c}. Anisotropic diffusion
is more pronounced for a higher moment of inertia, the highest value
being reached for dumbbells with all the mass concentrated at the
endpoints of the needle (see Fig.~\ref{r2figD2}). Lowering the
density leads to shrinking of the intermediate time regime, such
that for small densities ($\rho_0\leq 10$) one only observes a
crossover from ballistic motion to isotropic diffusion. For these
densities the agreement between theory and simulation is very good, as
shown in Fig.~\ref{r2figD6a}.

Since the intermediate anisotropic regime is only present for high
densities, we conclude that the Perrin equation is an adequate
description of the dynamics of thin hard rods only for high density.
In order to derive it from a microscopic theory, 
it is not sufficient to let the wavenumber become small;
in addition
the rotational diffusion constant has to be small or in other words
the rotational timescale has to be long.

A perspective for future work based on the microscopic theory presented here
would be to exploit the correlation function to calculate intermediate
scattering functions. These could then be measured in depolarized light
scattering experiments. Moreover, the microscopic theory is valid for all wave
vectors. However, up to now, a diagonalization procedure used to derive
explicit expressions for $\varphi_{ll'}$ exists only for small wave vectors.

\appendix

\section{Matrix elements}
In this appendix, the essential steps necessary to calculate the matrix
elements   $\chi_0$, $\Omega_0({\bf u'},{\bf u})$,
$\chi_1({\bf u'},{\bf u})$, and $\Omega_1({\bf u'},{\bf u})$ are presented.

\subsection{Generating functional}
The structure of the collision operator requires a specific coordinate system
for the translational momenta (or velocities).

In fact, the operator $i{\cal
L}^\prime_+$ appearing in matrix elements is reduced to the $N$-fold operator
for collision of needle $0$ with needle $1$,
\begin{align}
i{\cal L}^\prime_+ &=N\left|{\bf V}\cdot{\bf u}_\perp\right|
\theta(-{\bf V}\cdot{\bf u}_\perp \sigma)
\theta\left(\frac{L}{2}-|s_{01}|\right) 
\theta\left(\frac{L}{2}-|s_{10}|\right)
\delta\left(|{\bf r}_{01}^\perp|-0^+\right)
\left(b_{01}^+-1\right).
\end{align}
The equality
\beq
\dabl{}{t}|{\bf r}_{12}^\perp|={\bf u}_\perp\cdot{\bf V}_r
\mathrm{sgn}({\bf r}_{12}\cdot{\bf u}_\perp)
\eeq
has been used, and $\sigma$ is short for
$\mathrm{sgn}({\bf r}_{12}\cdot{\bf u}_\perp)$.
The relative velocity of contact points ${\bf V}$ and its
representation in the orthonormal basis ${\bf u}_\perp, {\bf
  u}_0^\perp, {\bf u}_0$
have been defined in
section~\ref{sec:coll}.  Before performing the transformation,
let us recall the definition of the integration measure in average
$\langle\dots\rangle$ in terms of canonical coordinates:
\beq
d\Gamma= \prod_{i=0}^1 d{\bf r}_i d\varphi_i d\theta_i
d{\bf p}_{{\bf r}_i}
dp_{\varphi_i}
dp_{\theta_i}
\eeq
The normalization reads as
\beq
Z^{-1}=\frac{1}{V^2}\frac{1}{(4\pi)^2}\frac{1}{(2\pi m T)^{3}}\frac{1}{(2\pi I T)^{2}}
\eeq
Next, the canonical coordinates are transformed to new ones which are standard
normally distributed (with respect to the Boltzmann factor $\exp(-\beta
{\cal H})$). Changing moreover translation momenta ${\bf p}_{{\bf r}_i}$ to
CMS momentum and relative momentum, the transformation reads as follows:
\beqa
\hbox{\bm $\chi$} &=& \frac{1}{\sqrt{2Tm}}({\bf p}_0-{\bf p}_1)\\
\hbox{\bm $\gamma$} &=& \frac{1}{\sqrt{2Tm}}({\bf p}_0+{\bf p}_1)\\
{\bf P}_{\theta_i} &=& \frac{1}{\sqrt{IT}}{\bf p}_{\theta_i}\\
{\bf P}_{\varphi_i} &=& \frac{1}{\sqrt{IT}}\frac{{\bf p}_{\varphi_i}}{\sin(\theta_i)}
\eeqa
with $i=0,1$. Positions are transformed according to
\beqa
{\bf r}_{01} &=& {\bf r}_0-{\bf r}_1\\
{\bf r} &=& {\bf r}_0
\eeqa
The resulting integration with respect to ${\bf r}_{01}$ will be split into the
following three components:
\beqa
c &=& {\bf r}_{01}\cdot{\bf u}_\perp\\
b=s_{10} &=& \frac{{\bf r}_{01}\cdot{\bf u}^\perp_0}{\sqrt{1-({\bf
      u}_0\cdot{\bf u}_1)^2}}\\
a=-s_{01} &=& {\bf r}_{01}\cdot{\bf u}_0-\frac{({\bf u}_0\cdot{\bf u}_1)({\bf r}_{01}\cdot{\bf u}^\perp_0)}{\sqrt{1-({\bf
      u}_0\cdot{\bf u}_1)^2}}
\eeqa
This decomposition complements the one for the relative velocity of contact
points. The corresponding Jacobian is given by
\beq
J=\frac{\partial(r_{01}^x,r_{01}^y,r_{01}^z)}{\partial(a,b,c)}=\sqrt{1-({\bf
      u}_0\cdot{\bf u}_1)^2}
\eeq
The relative velocity of contact points given in terms of the new coordinates
reads as follows:
\beqa
{\bf V}_r=\sqrt{\frac{2T}{m}}\hbox{\bm $\chi$}
&-&a\sqrt{\frac{T}{I}}(P_{\theta_0}\hat{\bf e}_{\theta_0}+
P_{\varphi_0}\hat{\bf e}_{\varphi_0})\nonumber\\
&-&b\sqrt{\frac{T}{I}}(P_{\theta_1}\hat{\bf e}_{\theta_1}+
P_{\varphi_1}\hat{\bf e}_{\varphi_1})
\eeqa
As the velocity ${\bf V}_r$ and relative position ${\bf r}_{01}$ have been
decomposed with respect to the base $({\bf u}_\perp,{\bf u}_0^\perp,{\bf
u}_0)$, the expression given just above will be rewritten likewise. Following a
suggestion by Huthmann, one may transform from standard normally distributed
variables
$(P_{\varphi_i}, P_{\theta_i})$ with respect to the base $(\hat{\bf
  e}_{\varphi_i}, \hat{\bf e}_{\theta_i})$ to new
standard normally distributed variables $(v_i, w_i)$ with respect to the base
$({\bf   u}^\perp_i, {\bf u}_\perp)$, giving the following expression for
${\bf V}_r$:
\beqa
\label{V2chi}
{\bf V}_r=\sqrt{\frac{2T}{m}}\hbox{\bm $\chi$}
&-&a\sqrt{\frac{T}{I}}(v_0{\bf u}^\perp_0+w_0{\bf u}_\perp)\nonumber\\
&-&b\sqrt{\frac{T}{I}}(v_1{\bf u}^\perp_1+w_1{\bf u}_\perp)
\eeqa
Later, on ${\bf u}^\perp_1$ will be eliminated using
\beq
{\bf u}^\perp_1=-\tilde{A} {\bf u}_0+\tilde{B}{\bf u}^\perp_0
\eeq
where
\beqa
\tilde{A} &=&\sqrt{1-({\bf u}_0\cdot{\bf u}_1)^2}\\
\tilde{B} &=& ({\bf u}_0\cdot{\bf u}_1)
\eeqa
So a decomposition of ${\bf V}_r$ with respect to the base $({\bf u}_\perp,{\bf
u}_0^\perp,{\bf u}_0)$ results.

In terms of the variables $(\hbox{\bm $\gamma$},\hbox{\bm $\chi$}, v_0,
w_0, v_1, w_1)$ the 2-needle Hamiltonian reads
\beq
\beta{\cal H}=\frac{1}{2}\left(
\hbox{\bm $\gamma$}^2+\hbox{\bm $\chi$}^2+ v_0^2+
w_0^2+ v_1^2+ w_1^2
\right).
\eeq
Accordingly,
the integration measure multiplied by the normalization changes to
\beq
Z^{-1}d\Gamma=\frac{1}{V^2}\frac{1}{(4\pi)^2}\frac{1}{(2\pi)^{5}}
d{\bf r}
dadbdc
d\Omega_0 d\Omega_1
d\hbox{\bm $\gamma$}
d\hbox{\bm $\chi$}
dv_0 dw_0
dv_1 dw_1
\eeq
where $d\Omega_i=d\varphi_i \sin(\theta_i)d\theta_i$ defines the integration
with respect to spatial angles.
Now, the collision operator involves a projection of ${\bf V}_r$. Moreover,
 $\hbox{\bm $\gamma$}$ is
related to ${\bf V}_r$ by Eq.~(\ref{V2chi}). Therefore, $\hbox{\bm $\gamma$}$
will have to be expressed in terms of ${\bf V}_r$ in the integration measure
and the Hamiltonian.
The integration measure involving a factor by $d\hbox{\bm
$\chi$}=\left(\frac{m}{2T}\right)^{3/2}dxdydz$, the Hamiltonian is
\begin{align}
\beta{\cal H}&= \frac{1}{2}\Bigg[
\frac{m}{2T}(x^2+y^2+z^2)
+(v_0^2+w_0^2)
\left(1+a^2\frac{m}{2I}\right)
\nonumber\\
&\quad +(v_1^2+w_1^2)
\left(1+b^2\frac{m}{2I}\right)
+
x\frac{m}{T}\sqrt{\frac{T}{I}}(aw_0+bw_1)\nonumber\\
&\quad +y\frac{m}{T}\sqrt{\frac{T}{I}}(\tilde{B}bv_1+av_0)
-z\frac{m}{T}\sqrt{\frac{T}{I}}(\tilde{A}bv_1)
\nonumber\\
&\quad+
ab\frac{m}{I}\tilde{B}v_0v_1
+ab\frac{m}{I}w_0w_1
+\hbox{\bm $\gamma$}^2\Bigg]
\end{align}
The variable $x$ will require a particular treatment as will be shown later on.
Moreover, the variable $\hbox{\bm $\gamma$}$, denoting the CMS momentum, will
not appear in matrix elements involving only relative translational momenta, and therefore will be treated seperately.
Defining the vector ${\bf X}$ by
\beq
{\bf X}^T=(y,z,v_0,w_0,v_1,w_1)
\eeq
the Hamiltonian may be rewritten in the following form
\beq
\beta{\cal H}=\frac{1}{2}{\bf X}^T{\bf A}{\bf X}
+\frac{1}{2}\left(
\frac{m}{2T}x^2
+x\frac{m}{T}\sqrt{\frac{T}{I}}(aw_0+bw_1)
\right) +\frac{1}{2}\hbox{\bm $\gamma$}^2
\eeq
where ${\bf A}$ is the following matrix:
\begin{widetext}
\beq
{\bf A}=\left(
\begin{array}{cccccc}
\frac{m}{2T} & 0 & \frac{m}{2T}\sqrt{\frac{T}{I}}a & 0 &
\frac{m}{2T}\sqrt{\frac{T}{I}}b\tilde{B} & 0\\
0 & \frac{m}{2T} & 0 & 0 & \frac{m}{2T}\sqrt{\frac{T}{I}}\tilde{A}b & 0\\
\frac{m}{2T}\sqrt{\frac{T}{I}}a & 0 & \left(1+a^2\frac{m}{2I}\right) &
0 &  ab\frac{m}{2I}\tilde{B} & 0 \\
0 & 0 & 0 & \left(1+a^2\frac{m}{2I}\right) & 0 & ab\frac{m}{2I} \\
\frac{m}{2T}\sqrt{\frac{T}{I}}b\tilde{B} &
\frac{m}{2T}\sqrt{\frac{T}{I}}\tilde{A}b & ab\frac{m}{2I}\tilde{B} & 0 &
\left(1+b^2\frac{m}{2I}\right) & 0 \\ 0 & 0 & 0 & ab\frac{m}{2I} & 0 &
\left(1+b^2\frac{m}{2I}\right) \end{array}
\right)
\eeq
Let us now consider the generating functional with respect to
\beq
{\bf H}^T=(h_y,h_z,h_{v_0},h_{w_0},h_{v_1},h_{w_1}),
\eeq
namely
\begin{align}
G({\bf H}) &= \int d{\bf X}\exp\left(
-\beta{\cal H}+{\bf H}^T{\bf X}
\right)\nonumber\\
 &= \int d{\bf X}\exp\left(
-\frac{1}{2}{\bf X}^T{\bf A}{\bf X}
-\frac{1}{2}\left(
\frac{m}{2T}x^2
+x\frac{m}{T}\sqrt{\frac{T}{I}}(aw_0+bw_1)
\right)+{\bf H}^T{\bf X}
\right)\nonumber\\
\end{align}
The integration is gaussian and the result is:
\beq
G({\bf
  H})=(2\pi)^3\frac{2T}{m}\left(1+\frac{m}{2I}(a^2+b^2)\right)^{-1/2}
\exp\left(-\beta{\cal H}_{\text{eff}}\right)
\eeq
with the effective Hamiltonian
\beqa
\beta{\cal H}_{\text{eff}} &=&
\frac{x^2}{2}\left(
\frac{m}{2T}\right)
\frac{1}{\left(1+\frac{m}{2I}(a^2+b^2)\right)}
-\frac{1}{2}{\bf H}^T{\bf A}^{-1}{\bf H}+\frac{1}{2}\hbox{\bm $\gamma$}^2
\nonumber\\
&+&x\frac{m}{2T\left(1+\frac{m}{2I}(a^2+b^2)\right)}
\sqrt{\frac{T}{I}}
\left(
ah_{w_0}
+bh_{w_1}
\right)\nonumber\\
\eeqa
The inverse matrix ${\bf A}^{-1}$ is given by
\beqa
% &&{\bf A}^{-1}=
% \nonumber\\
% &&
\left(
\begin{array}{cccccc}
\frac{2T}{m}\left(1+\frac{m}{2I}(a^2+b^2\tilde{B}^2)\right) & -\frac{T}{I}(\tilde{A}\tilde{B}b^2) & -\sqrt{\frac{T}{I}}a & 0 &
-\sqrt{\frac{T}{I}}b\tilde{B} & 0\\
-\frac{T}{I}(\tilde{A}\tilde{B}b^2) &  \frac{2T}{m}\left(1+\frac{m}{2I}b^2\tilde{A}^2\right) &
0 & 0 & \sqrt{\frac{T}{I}}\tilde{A}b & 0\\
-\sqrt{\frac{T}{I}}a & 0 & 1 &
0 &  0 & 0 \\
0 & 0 & 0 & \frac{1+b^2\frac{m}{2I}}{1+\frac{m}{2I}(a^2+b^2)} & 0 & -\frac{ab\frac{m}{2I}}{1+\frac{m}{2I}(a^2+b^2)} \\
-\sqrt{\frac{T}{I}}b\tilde{B} & \sqrt{\frac{T}{I}}\tilde{A}b & 0 & 0 & 1 & 0 \\
0 & 0 & 0 & -\frac{ab\frac{m}{2I}}{1+\frac{m}{2I}(a^2+b^2)} & 0 &
\frac{1+a^2\frac{m}{2I}}{1+\frac{m}{2I}(a^2+b^2)}
\end{array}
\right)\nonumber\\
\eeqa
Whenever matrix elements involving the collision operator appear,
the delta function $\delta\left(|{\bf
r}_{01}^\perp|-0^+\right)=\delta\left(|c|-0^+\right)$ in terms of
new position coordinates as well as the dependance on $\mathrm{sgn}(c)$ need to be taken care
of. In this case the following relation is useful:
\begin{align}
\begin{split}
\int_{-\infty}^\infty dc\, \delta\left(|c|-0^+\right)\dots
f(\dots,\mathrm{sgn}(c)) 
&=\lim_{\epsilon\rightarrow 0^+}\int_{-\infty}^\infty dc
\left(\delta(c+\epsilon)+\delta(c-\epsilon)\right)\dots
f(\dots,\mathrm{sgn}(c))\\
&=\dots(f(\dots,+1)+f(\dots,-1))
\end{split}
\end{align}
\end{widetext}

\subsection{Matrix elements}
Matrix elements explicitly involving the collision operator can be easily
calculated when represented as moments of the variables
$(x,y,z,v_0,w_0,v_1,w_1,\hbox{\bm $\gamma$})$ using the generating functional
derived above. In fact moments involving $(y,z,v_0,w_0,v_1,w_1)$ can be
obtained by replacing expression involving a variable say $y$ to some power by
the respective derivative with the order equaling the power. The integrations
with respect to $x$ and $ \hbox{\bm $\gamma$}$ are treated separately.

In fact, a matrix element $\langle A^*i{\cal L}^\prime_+ A\rangle$
usually has the following form
\begin{align}
\langle A^* i{\cal L}A \rangle
&= C
\int_{-L/2}^{L/2}da
\int_{-L/2}^{L/2}db\nonumber\\
&\quad\times\int\dots\int_{\Omega_0,\Omega_1
,x
,y
,z , v_0, w_0
,v_1, w_1,\hbox{\bm $\gamma$}}
J (\Omega_0,\Omega_1)
e^{-\beta {\cal H}}\nonumber\\
&\quad\times
\left(\Theta(-x)+\Theta(x)\right)
|x|\nonumber\\
&\quad\times
f(\Omega_0,\Omega_1,x
,y
,z , v_0, w_0
,v_1, w_1,\hbox{\bm $\gamma$})
\end{align}
where
\beq
C=\frac{1}{Z}VN(m/2T)^{3/2}
\eeq
The function $f(\Omega_0,\Omega_1,x
,y
,z , v_0, w_0
,v_1, w_1,\hbox{\bm $\gamma$})$ contains the functional form of the factors
$A^*$ and $A$ in the matrix element after $\left(b_{01}^+-1\right)$ has been
carried out. Other factors arising from the collision operator have been
written separately. The factor $C$ is a normalization factor which results
after integrating with respect to the CMS position ${\bf r}$ and transforming
form $\hbox{\bm $\chi$}$ to $(x,y,z)$. The generating functional now permits
reducing the number of integrations as follows:
\begin{widetext}
\begin{align}
\langle A^* i{\cal L}A \rangle
&= C
\int_{-L/2}^{L/2}da
\int_{-L/2}^{L/2}db
\int\dots\int_{\Omega_0,\Omega_1
,x
,\hbox{\bm $\gamma$}}
J(\Omega_0,\Omega_1)
\left(\Theta(-x)+\Theta(x)\right)
|x|
\nonumber\\
&\quad\times f(\Omega_0,\Omega_1,x,\hbox{\bm $\gamma$}
,\partial_{h_y},
,\partial_{h_z} , \partial_{h_{v_0}},
\partial_{h_{w_0}},
\partial_{h_{v_1}},
\partial_{h_{w_1}})
G({\bf H})
\end{align}
In our case, no dependances on  $\hbox{\bm
  $\gamma$}$ exist, so
the corresponding integration may be immediately carried out, giving a factor
of $(2\pi)^{3/2}$.

Let us outline the calculation for the matrix element
$\Omega_1({\bf u'},{\bf u})$. It involves 2 matrix elements $\langle B_i^* {\cal L}B_i
\rangle$ for $i=1,2$ resulting from the two terms in $\dot{A}_{\bf u}$. Taking
into account the change of $\Delta\dot{\bf u}_0$ and $\Delta{\bf p}$ after a
collision (which are given for the general case by Eq.s(\ref{p.rough}) and
(\ref{udot.rough})), the matrix element reads as follows:
\begin{align}
\langle B_1^* {\cal L}B_1 \rangle
&= (-i)C
\int_{-L/2}^{L/2}da
\int_{-L/2}^{L/2}db
\int\dots\int_{\Omega_0,\Omega_1
,x
,y
,z , v_0, w_0
,v_1, w_1}
J
\nonumber\\
&\quad\times
e^{-\beta {\cal H}}
\left(
 {\bf k}\cdot\frac{1}{2}\left({\bf V}_r
+a\sqrt{\frac{T}{I}}(v_0{\bf u}^\perp_0+w_0{\bf u}_\perp)
+b\sqrt{\frac{T}{I}}(v_1{\bf u}^\perp_1+w_1{\bf u}_\perp)
\right)\right)\nonumber\\
&\quad\times|x|
\left(\theta(-x)+\theta(x)\right)
\frac{1}{m}{\bf k}\cdot\left(
\alpha{\bf u}_\perp+\gamma_1{\bf u}_0+\gamma_2{\bf u}^\perp_0
\right)
\delta({\bf u'}-{\bf u}_0)\delta({\bf u}-{\bf u}_0)
\end{align}
and
\begin{align}
\langle B_2^* {\cal L}B_2 \rangle
&= (-i)\nabla_{\bf u'}^\mu\nabla_{\bf u}^\nu
C
\int_{-L/2}^{L/2}da
\int_{-L/2}^{L/2}db
\int\dots\int_{\Omega_0,\Omega_1
,x
,y
,z , v_0, w_0
,v_1, w_1}
J
\nonumber\\
&\quad\times
\frac{am}{2I}
e^{-\beta {\cal H}}
\left(
{\bf u}_0({\bf u}_0\cdot{\bf V}_r)+
{\bf u}_0({\bf u}_0\cdot{\bf u}^\perp_1)b\sqrt{\frac{T}{I}}v_1
-{\bf V}_r
-a\sqrt{\frac{T}{I}}(v_0{\bf u}^\perp_0+w_0{\bf u}_\perp)
-b\sqrt{\frac{T}{I}}(v_1{\bf u}^\perp_1+w_1{\bf u}_\perp)
\right)_\mu
\nonumber\\
&\quad\times|x|
\left(\theta(-x)+\theta(x)\right)
\left(-\frac{a}{I}\right)
\left(\alpha{\bf u}_\perp+\gamma_2 {\bf u}^\perp_0
\right)_\nu
\delta({\bf u'}-{\bf u}_0)\delta({\bf u}-{\bf u}_0)
\end{align}
The coefficients $\alpha$, $\gamma_1$, and $\gamma_2$ are defined in Eq.s
(\ref{alpha}) and (\ref{gamma1u2}), and contain dependancies on $x$, $y$, $z$,
$({\bf u}_0\cdot{\bf u}_1)$, as well as $a$ and $b$.

Using the generating functional as above one finds the following results,
\begin{align}
\langle B_1^* {\cal L}B_1 \rangle&=
in_0 \frac{1}{8\pi^{5/2}}\left(\frac{T}{m}\right)^{3/2}L^2
\left[
c_\parallel ({\bf k}\cdot {\bf u})^2 + c_\perp (k^2-({\bf k}\cdot {\bf u})^2)
\right]\delta({\bf u'}-{\bf u})\\
\langle B_2^* {\cal L}B_2 \rangle&=
in_0 \frac{1}{8\pi^{5/2}}\left(\frac{T}{m}\right)^{3/2}L^4
\left(
\frac{m^2}{I^2}
\right)c_R\hat{L}^2
\delta({\bf u'}-{\bf u})
\end{align}
\end{widetext}
The coefficients $c_\parallel$, $c_\perp$, and $c_R$ contain the remaining
integrations over $a$, $b$, and the relative angle between $\Omega_0$ and
$\Omega_1$, and are defined in appendix B. Adding the two matrix elements gives
the expression for  $\Omega_1({\bf u'},{\bf u})$ in Eq.~(\ref{omega1}).

\section{Some geometrical integrals}
\label{geomfactors}

Let us first define
\beq
I_i=\int_{-1/2}^{1/2}da \int_{-1/2}^{1/2}db \frac{a^i}{\sqrt{1+\D(a^2+b^2)}}
\eeq
The constants appearing in Eq.(\ref{omega1}) are given explicitly
as follows where $\D=mL^2/(2I)$
\begin{align}
c_\parallel&=\left(\frac{1+\beta}{2}\right)
2\pi \int_{-1/2}^{1/2}da\, \int_{-1/2}^{1/2}db\,
\left(1+\D(a^2+b^2)\right)^{1/2}\nonumber\\
&\quad\times
\int_{-1}^1 dx\, (1-x^2)^{1/2}
\frac{(1+\D a^2+\D b^2x^2)}{\left(1+\D(a^2+b^2)+\D^2a^2b^2(1-x^2)\right)}\\
c_\perp &=\pi^2 I_0 +
\left(\frac{1+\beta}{2}\right)\pi\int_{-1/2}^{1/2}da\, \int_{-1/2}^{1/2}db \,
\left(1+\D(a^2+b^2)\right)^{1/2}\nonumber\\
&\quad\times\int_{-1}^1 dx\, (1-x^2)^{1/2}
\frac{\left(1+\D 
b^2(1-x^2)\right)}{\left(1+\D(a^2+b^2)+\D^2a^2b^2(1-x^2)\right)}\\
c_R &=\pi^2 I_2+
\left(\frac{1+\beta}{2}\right)\pi \int_{-1/2}^{1/2}da\, \int_{-1/2}^{1/2}db\,
a^2
\left(1+\D(a^2+b^2)\right)^{1/2}\nonumber\\
&\quad\times\int_{-1}^1 dx\, (1-x^2)^{1/2}
\frac{\left(1+\D 
b^2(1-x^2)\right)}{\left(1+\D(a^2+b^2)+\D^2a^2b^2(1-x^2)\right)}
\end{align}
They have been evaluated numerically for $\D=6$ and $\D=2$ (see
Table~\ref{ctab}).

\section{Some properties of spheroidal wave functions}
\label{spheroidal}
For a general introduction to spheroidal wave functions we refer the reader to
Refs.~\onlinecite{stratton,flammer:57}.  In this work we only need spheroidal functions of
either prolate or oblate type. The prolate functions are defined as the
solutions of the eigenvalue equation
\begin{align}
\label{spheroiddef}
\left(\frac{d}{d\eta}(1-\eta^2)\frac{d}{d\eta}+
\lambda_{lm}(\zeta)-\frac{m^2}{1-\eta^2}-\zeta^2\eta^2
\right)S_{lm}(\zeta,\eta)&=0,
\end{align}
where $l\ge 0$ and $m$ are integers obeying $-l \le m \le l$ and
$\zeta$ is a real constant.  The oblate functions can be obtained from
$S_{lm}(\zeta,\eta)$ by replacing $\zeta$ by $i\zeta$. We prefer 
working with an orthonormal basis and therefore choose the
normalization
\begin{align}
  \int d\eta\,S_{lm}^*(\zeta,\eta)S_{l'm}(\zeta,\eta)&=\delta_{ll'}
\end{align}
which is unfortunately different from the one used in
Refs.~\onlinecite{stratton,flammer:57}. 

The $S_{lm}(\zeta,\eta)$ can be expressed in terms of associated Legendre
functions $P_{lm}$ as
\begin{align}
  S_{lm}(\zeta,\eta) &= \frac{1}{N_{lm}(\zeta)}{\sum_{r}}'d_r^{lm}(\zeta)
P_{r+m,m}(\eta).
\end{align}
Here $d_r^{lm}(\zeta)$ are expansion coefficients and 
\begin{align}
  N_{lm}(\zeta) &= \sqrt{{\sum_r}'(d_r^{lm}(\zeta))^2\frac{2}{2r+2m+1}
\frac{(r+2m)!}{r!}}
\end{align}
is the normalization constant. For methods to obtain
$\lambda_{lm}(\zeta)$ and $d_r^{lm}$ from Eq.~\eqref{spheroiddef} we
again refer the reader to Refs.~\onlinecite{stratton,flammer:57}. For our purposes it
is sufficient to know the eigenvalues $\lambda_{lm}(\zeta)$ up to
order $\zeta^2$ and the scalar product
\begin{align}
  (S_{lm}(\zeta), S_{l'm}(0)) &\equiv \int d\eta\,
  S_{lm}^*(\zeta,\eta)S_{l'm}(0,\eta)
\end{align}
up to order $\zeta^4$. The results of this tedious calculation (which is
omitted here for the sake of brevity) are
\begin{align}
  \lambda_{lm}(\zeta) &= \lambda_{lm}^{(0)} + \zeta^2\lambda_{lm}^{(2)} +
\mathcal{O}(\zeta^4) \\
&= l(l+1) + \zeta^2  
\frac{2l(l+1)-2m^2-1}{(2l+3)(2l-1)} + \mathcal{O}(\zeta^4)
\end{align}
and
\begin{align}
(S_{lm}(\zeta), S_{l'm}(0)) &= \delta_{ll'} + \zeta^2 g_{mll'} +
\zeta^4 h_{mll'} + \mathcal{O}(\zeta^6)
\end{align}
where
\begin{align}
\begin{split}
	g_{mll'} &= \delta_{l,l'+2}
	\frac{\sqrt{(l-1-m)(l-1+m)(l-m)(l+m)}}
	{2(2l-1)^2\sqrt{(2l-3)(2l+1)}}- \\
	&\quad\delta_{l+2,l'}\frac{\sqrt{(l'-1-m)(l'-1+m)(l'-m)(l'+m)}}
	{2(2l'-1)^2\sqrt{(2l'-3)(2l'+1)}} 
\end{split}\\
\begin{split}
	h_{mll'} &= \frac{1-\delta_{ll'}}{\lambda_{lm}^{(0)}-\lambda_{l'm}^{(0)}}
	(a_{l'm}g_{ml,l'-2} + a_{l'+2,m}g_{ml,l'+2})- \\
	&\quad\frac{\lambda_{lm}^{(2)}-\lambda_{l'm}^{(2)}}
	{\lambda_{lm}^{(0)}-\lambda_{l'm}^{(0)}}g_{mll'}-
	\frac 12\delta_{ll'}(g_{ml,l+2}^2+g_{ml,l-2}^2)\quad\text{and}
\end{split}\\
a_{lm} &= \frac{\sqrt{(l-1+m)(l-1-m)(l+m)(l-m)}}
{(2l-1)\sqrt{(2l+1)(2l-3)}}.
\end{align}
Here it is implied that all coefficients are 0 if their indices lie
outside their range of validity $l,l'\ge 0$ and $-l, -l'\le m\le
l,l'$. For some values of the indices the coefficients appear to have
vanishing denominators. However, in these cases the prefactors are 0,
too, and the respective terms should be taken to be 0.

\section{Diagonalization of $\Phi$}
\label{diag}
In order to diagonalize the matrix of correlation functions
$\varphi_{\eta\eta'}(\mathbf{k}, z)$, it is sufficient to diagonalize
$B_{\eta\eta'}=z(\chi^{-1})_{\eta\eta'} +
(\chi^{-1}\Omega\chi^{-1})_{\eta\eta'}$.  Since $\chi_{\eta\eta'}$ is diagonal
in the basis of (normalized) Legendre polynomials $\sqrt{l+\frac
  12}P_l(\eta)=S_{l0}(0,\eta)$ and $\Omega_{\eta\eta'}$ is diagonal in the
basis of $S_{l0}(ic,\eta)$, where $c^2=(c_\perp-c_\parallel)L^2k^2/(4\D^2c_R)$
is small in the limit $k^2\to 0$, we aim at a perturbative solution. The
starting point is the representation of $B$ in the basis $\{S_{l0}(0,\eta)\}$.
The matrix elements are
\begin{align}
  B_{ll'} &= \int d\eta\,d\eta'\,S_{l0}(0,\eta)B_{\eta\eta'}S_{l'0}(0,\eta').
\end{align}
Since $\chi$ is already diagonal in this basis, nothing needs to be done for
the term $z\chi^{-1}$ and we get $z(\chi^{-1})_{ll'}=z\delta_{ll'}/\chi_l$.

The term $\chi^{-1}\Omega\chi^{-1}$ is more complicated.  Inserting
$\mathbf{1}=\sum_l |S_{l0}(ic))(S_{l0}(ic)|$ in the two
places between $\chi^{-1}$ and $\Omega$ we get
\begin{align}
  B_{ll'} &= z\delta_{ll'}/\chi_l + (\chi^{-1}\Omega\chi^{-1})_{ll'} \\
  &=z\delta_{ll'}/\chi_l +
  \delta_{ll'}\frac{\Omega_{l}}{\chi_{l}^2} +
  \Delta B^{1}_{ll'} + \Delta B^{2}_{ll'} + \Delta B^{3}_{ll'},
  \quad \text{where}\\
  \Delta B^{1}_{ll'} &=
  c^2 g_{0ll'}\frac{\Omega_{l}-\Omega_{l'}}
  {\chi_{l}\chi_{l'}}, \\
  \Delta B^{2}_{ll'} &=
  \frac{c^4}{\chi_{0}}(h_{0l'0}\frac{\Omega_{l'}}{\chi_{l'}}
  \delta_{l0}+
  h_{0l0}\frac{\Omega_{l}}{\chi_{l}}\delta_{l'0}), \\
  \Delta B^{3}_{ll'} &= \frac{c^4}{\chi_{l}\chi_{l'}}
  \sum_{j}g_{0jl}g_{0jl'}\Omega_{j}
\end{align}
to order $k^2$.  The calculation is straightforward but care has to be
taken with $\chi_{0}$ and $\Omega_{0}$ as they are of order $k^2$
while all other eigenvalues of $\chi$ and $\Omega$ are of order 1.
Therefore the matrix elements $\Delta B^1_{02}$ and $\Delta B^1_{20}$
are of order $1$.  However, all other off-diagonal matrix elements of
$B$ are of order $k^2$ or smaller. The diagonal element $B_{00}$ is of
order $1/k^2$ while all other diagonal elements are of order $1$.  In
order to be able to use perturbation theory we employ a Jacobi
transformation
\begin{align}
  R_{ll'} &= \delta_{ll'} + 
  (\cos\frac{B_{02}}{B_{00}}-1)(\delta_{l0}\delta_{l'0} +
  \delta_{l2}\delta_{l'2}) + 
  \sin\frac{B_{02}}{B_{00}}(\delta_{l0}\delta_{l'2} - 
  \delta_{l2}\delta_{l'0}).
\end{align}
which makes $R B R^T = D + k^2 \Delta$ where $D$ is a diagonal matrix
and $\Delta$ is a perturbation. Note that $R$ is really only required
up to order $k^2$ which is
\begin{align}
R_{ll'} &= \delta_{ll'} + \frac{B_{02}}{B_{00}}(\delta_{l0}\delta_{l'2} - 
  \delta_{l2}\delta_{l'0}) + \mathcal{O}(k^4).
\end{align}
The eigenvalues $b_l$ of $B$ (up to order $k^2$) are given by the diagonal
entries of $RBR^T$ since the Jacobi transformation merely changes the basis
vectors but not the eigenvalues. The diagonal entries of $RBR^T$ are equal to
$b_l=B_{ll}=z/\chi_l+\Omega_l/\chi_l^2$ for all $l\ne0,2$. For the two
exceptions one gets $b_0=B_{00}+2B_{02}^2/B_{00}$ ($l=0$) and
$b_2=B_{22}-B_{02}^2/B_{00}$ ($l=2$). These are, however, higher order
corrections which are irrelevant for our purposes and we may therefore use
$b_l=z/\chi_l+\Omega_l/\chi_l^2$ for all $l$.

Finally, the eigenfunctions $S'_l(\eta)$ can be obtained through
perturbation theory. The matrix $B_{ll'}$ is symmetric but not
normal (i.e.\ $[B,B^\dagger]\ne0$, except if $z$ happens
to be purely imaginary).  Therefore $B$ and of course also $RBR^T$ do
not have an orthogonal system of eigenvectors and we have to calculate
both left and right eigenvectors perturbatively. This procedure yields
for the matrix $O^{\mathrm{r}}$ of right eigenvectors of $RBR^T$
($O^{\mathrm{r}}_{ll'}$ is the $l'$th component of the $l$th
eigenvector),
\begin{align}
O^{\mathrm{r}}_{ll'} &= \delta_{ll'} + (1-\delta_{ll'})
\frac{(RBR^T)_{l'l}}{b_l-b_{l'}},
\end{align}
while the matrix $O^{\mathrm{l}}$ of left eigenvectors is
\begin{align}
O^{\mathrm{l}}_{ll'} &= \delta_{ll'} + (1-\delta_{ll'})
\frac{(RBR^T)^*_{l'l}}{(b_l-b_{l'})^*} = (O^\mathrm{r}_{ll'})^*
\end{align}
and the asterisk denotes the complex conjugate.

The right and left eigenfunctions $S^{\mathrm{r}}_l(\eta)$ and
$S^{\mathrm{l}}_l(\eta)$ of the original operator $B$ are then
\begin{align}
S^{\mathrm{r}}_l(\eta) &= \sum_{l'}({O^{\mathrm{r}}}^*R)_{ll'}S_{l'0}(0,\eta)\\
S^{\mathrm{l}}_l(\eta) &= \sum_{l'}(O^{\mathrm{l}} R)^*_{ll'}S_{l'0}(0,\eta).
\end{align}
The product $O^{\mathrm{l}}R$ is, up to order $k^2$,
\begin{align}
(O^{\mathrm{l}}R)_{ll'} =
\delta_{ll'} +\frac{B_{02}}{B_{00}}(\delta_{l0}\delta_{l'2} - 
  \delta_{l2}\delta_{l'0}) + \\
(1-\delta_{ll'})
\frac{(RBR^T)^*_{l'l}}{(b_l-b_{l'})^*}
\label{ORprod}
\end{align}

As an application which is needed in the main text we calculate some
matrix elements of the correlation function $\varphi$ in the basis of
spherical functions $S_{l0}(0,\eta)$.  We have
\begin{align}
  \varphi_{ll'} &= -\sum_m (S_{l0}(0)|S^\mathrm{r}_m)
\frac{1}{z-\frac{4\pi}{b_m}}
(S^\mathrm{l}_m|S_{l'0}(0)) \\
&= -\sum_m ({O^{\mathrm{r}}}^*R)_{ml}\frac{1}{z-\frac{4\pi}{b_m}}
 (O^{\mathrm{l}} R)_{ml'},
\end{align}
and from this and Eq.~\eqref{ORprod} we get
\begin{align}
  \varphi_{00} &= -\frac{1}{z-\frac{4\pi}{b_0}} =
  -\frac{z+\Omega_0/\chi_0}{z (z+\Omega_0/\chi_0) - 4\pi\chi_0},
\end{align}
omitting terms of higher order than $k^2$. Likewise we obtain
\begin{align}
  \varphi_{20} &= - \frac{B_{02}}{B_{00}}\frac{1}{z-\frac{4\pi}{b_0}}
+\frac{B_{02}}{B_{00}}\frac{1}{z-\frac{4\pi}{b_2}} \\
&= 4\pi\frac{c^2}{9\sqrt{5}}\frac{-\Omega_2}
{\left(z^2+z\chi_0^{-1}\Omega_0-4\pi\chi_0\right)
\left(z^2+z\chi_2^{-1}\Omega_2-4\pi\chi_2\right)}.
\end{align}

%\section{The hierarchy of correlation functions $\varphi_{ll'}$}
%\label{hierarchy}

%\begin{references}
\bibliography{literatur}

\clearpage
\noindent\figcaption{\label{2rods}The collision plane $E_{01}$ spanned by the orientations of two rods.\hfill}

\noindent\figcaption{\label{phiuufig}Orientational correlation function 
$\varphi_{11}(\omega)$ from theory (Eq.~\eqref{phiuueq}) and
simulation for various densities $\rho_0$.} 

\noindent\figcaption{\label{r2figD6a}Mean square displacements $r_\parallel^2$ and
$r_\perp^2$ for homogeneous rods ($\D=6$) at densities $\rho_0=4$ and $\rho_0=10$ from theory and simulations. The uppermost line indicates the
exact ballistic behavior for small times. Distances are
measured in dimensionless units with $L=1$.}

\figcaption{\label{r2figD6b}Mean square displacements $r_\parallel^2$ and
$r_\perp^2$ for homogeneous rods ($\D=6$) at densities $\rho_0=40$ (simulation) and $\rho_0=100$ (theory). At these densities, an anisotropy window opens up between the parallel and perpendicular mean-square displacements.}

\figcaption{\label{r2figD6c}Mean square displacements $r_\parallel^2$ and
$r_\perp^2$ for homogeneous rods ($\D=6$) at densities $\rho_0=40$ and $\rho_0=100$.}

\figcaption{\label{r2figD2}Mean square displacements for dumbbell molecules
with $\D=2$.\hfill}

\clearpage
\noindent
\includegraphics[width=0.5\textwidth]{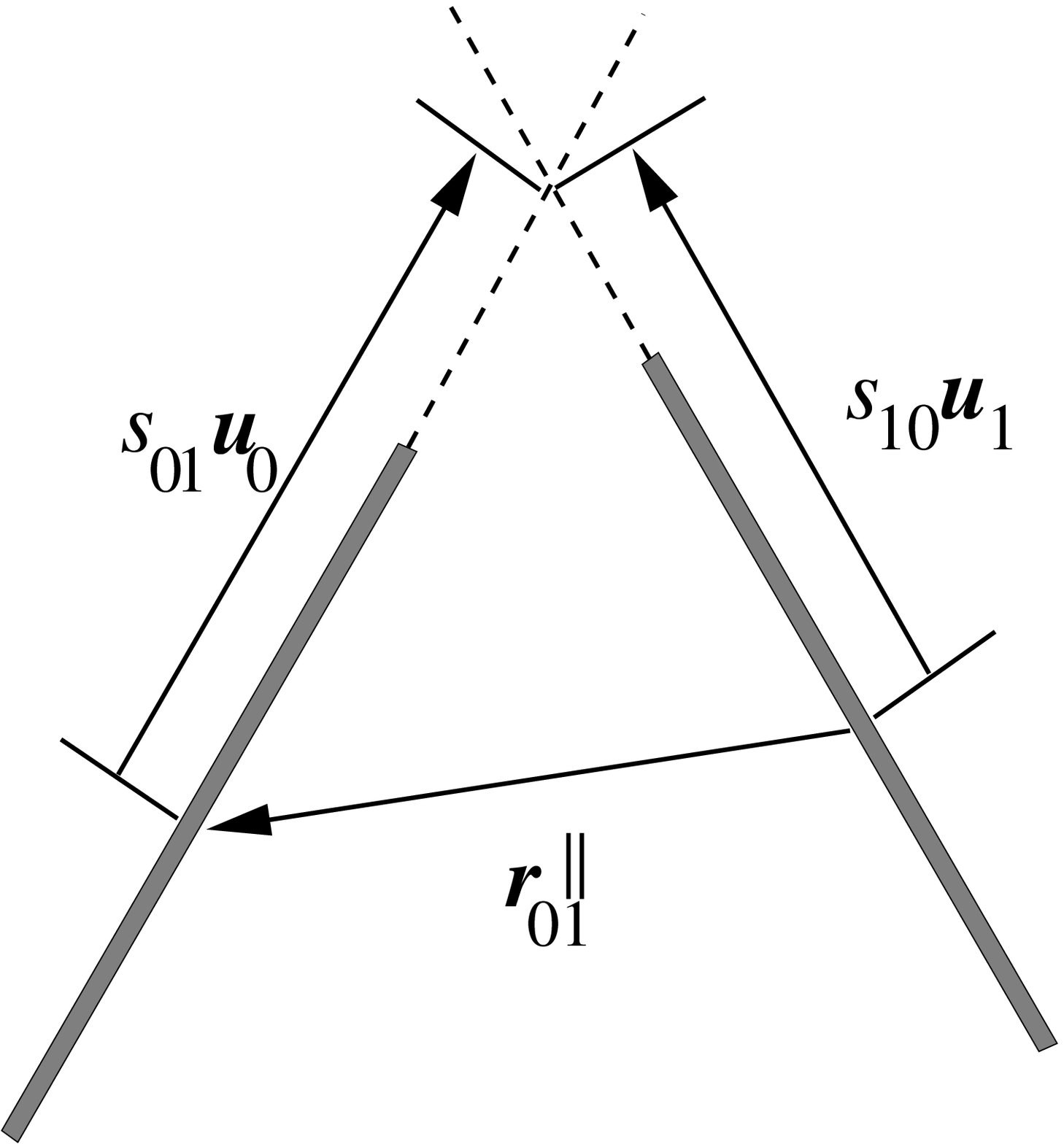}

\vfill
FIG.~\ref{2rods}
\thispagestyle{empty}

\clearpage
\noindent
\includegraphics[width=0.5\textwidth]{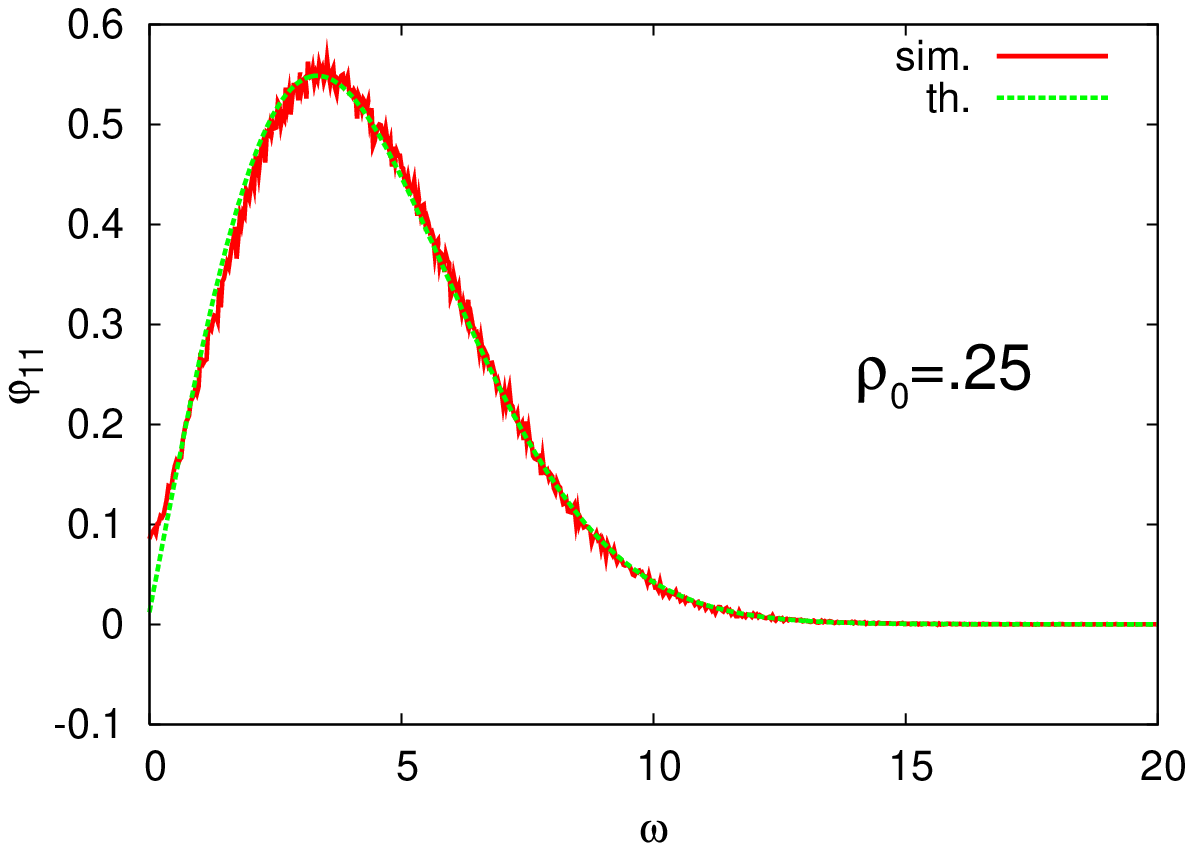}\hfill
\includegraphics[width=0.5\textwidth]{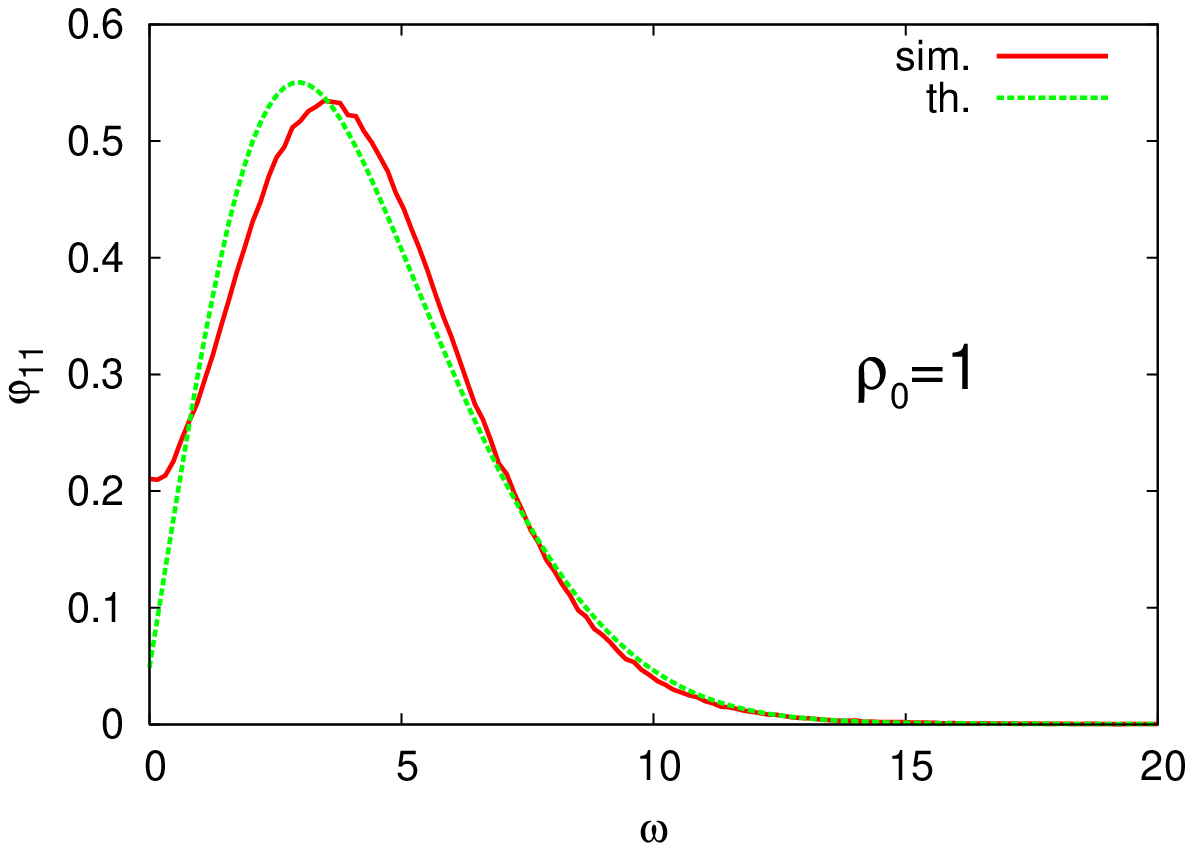}\\
\includegraphics[width=0.5\textwidth]{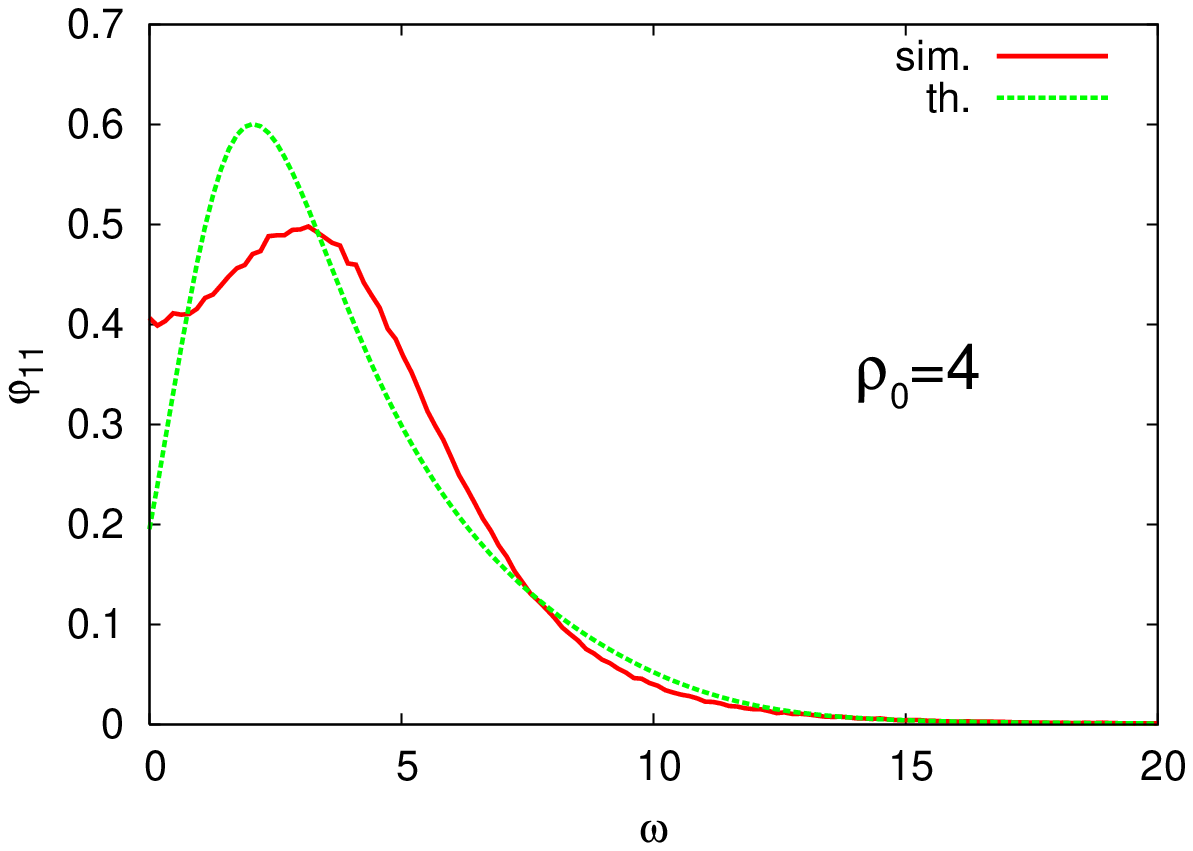}\hfill
\includegraphics[width=0.5\textwidth]{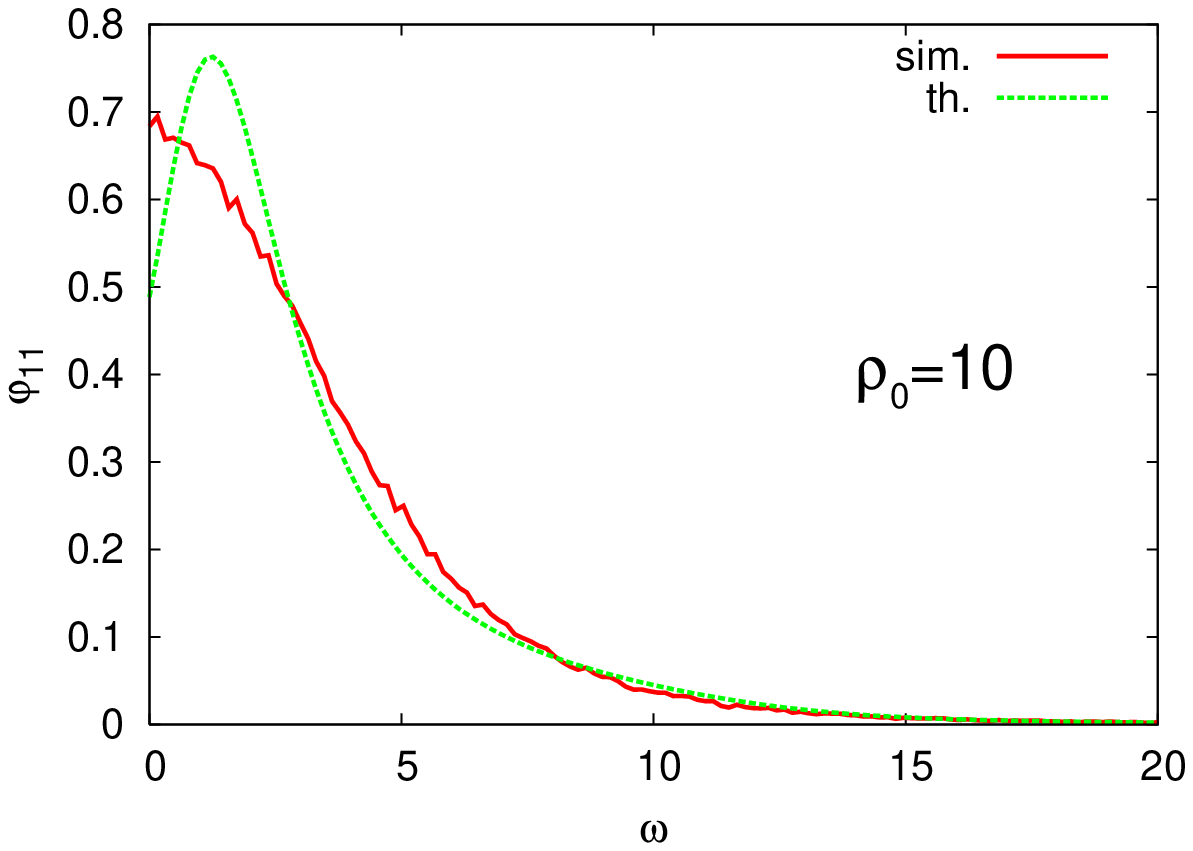}\\
\includegraphics[width=0.5\textwidth]{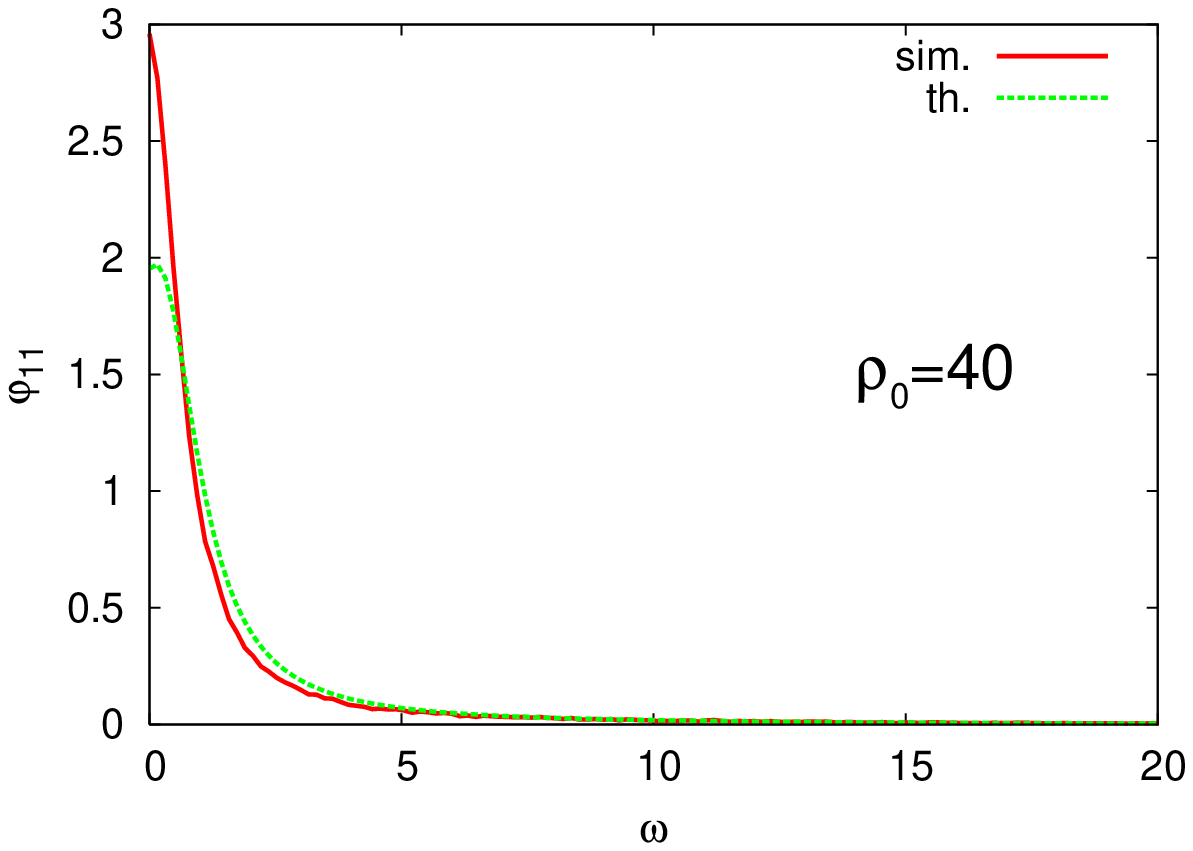}\\

\vfill
FIG.~\ref{phiuufig}
\thispagestyle{empty}

\clearpage
\noindent
\includegraphics[width=0.8\textwidth]{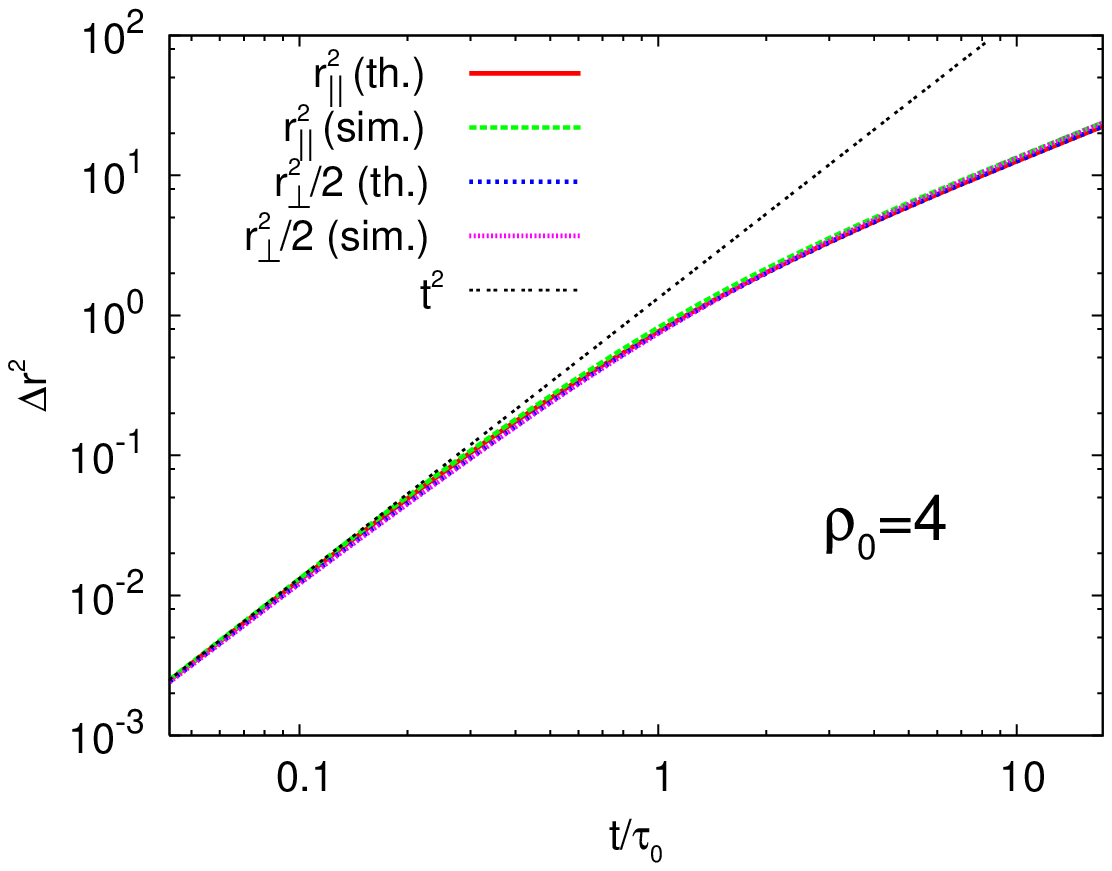}\\
\includegraphics[width=0.8\textwidth]{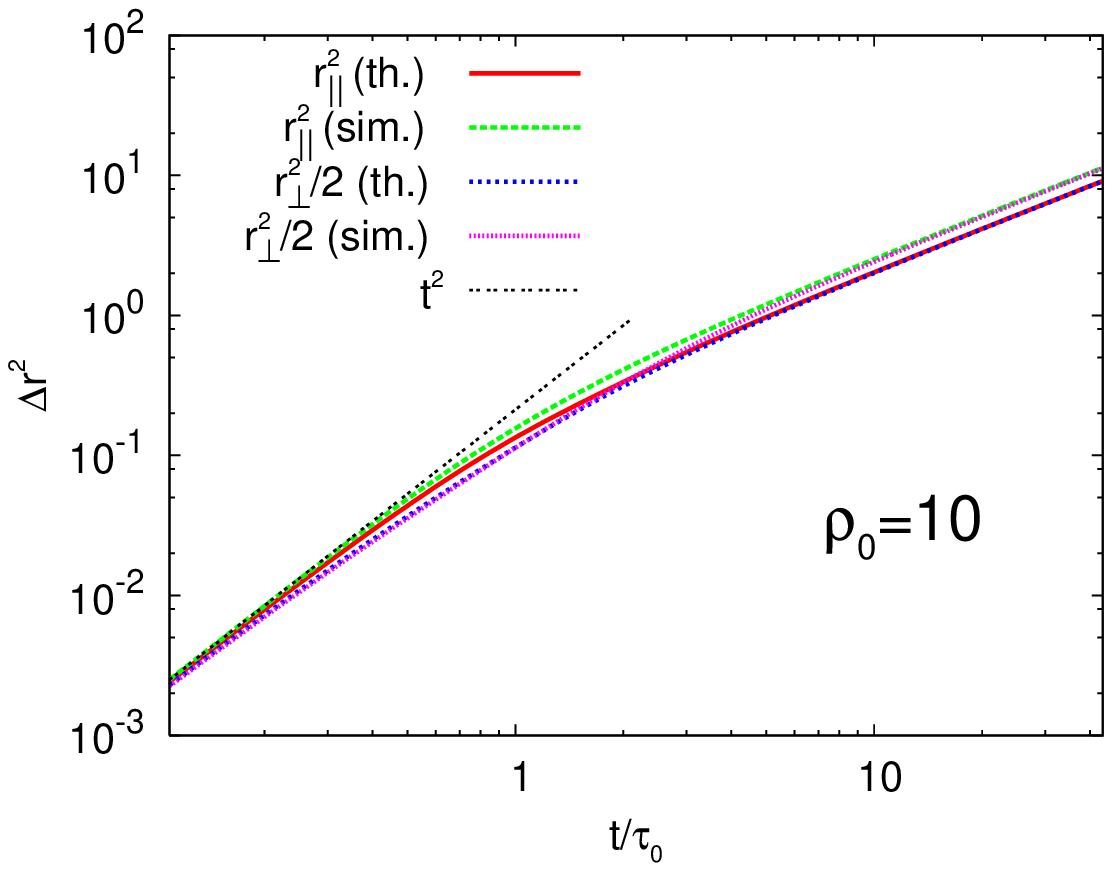}

\vfill
FIG.~\ref{r2figD6a}
\thispagestyle{empty}

\clearpage
\noindent
\includegraphics[width=0.8\textwidth]{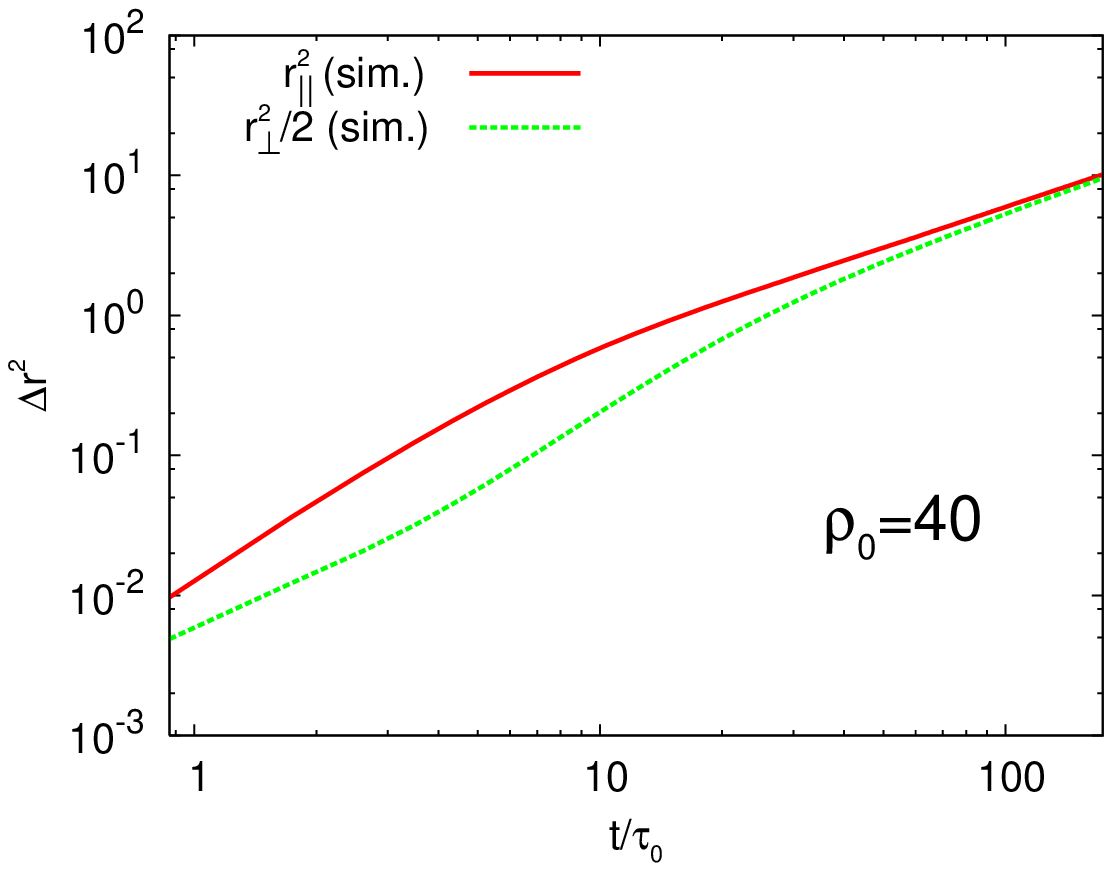}\\
\includegraphics[width=0.8\textwidth]{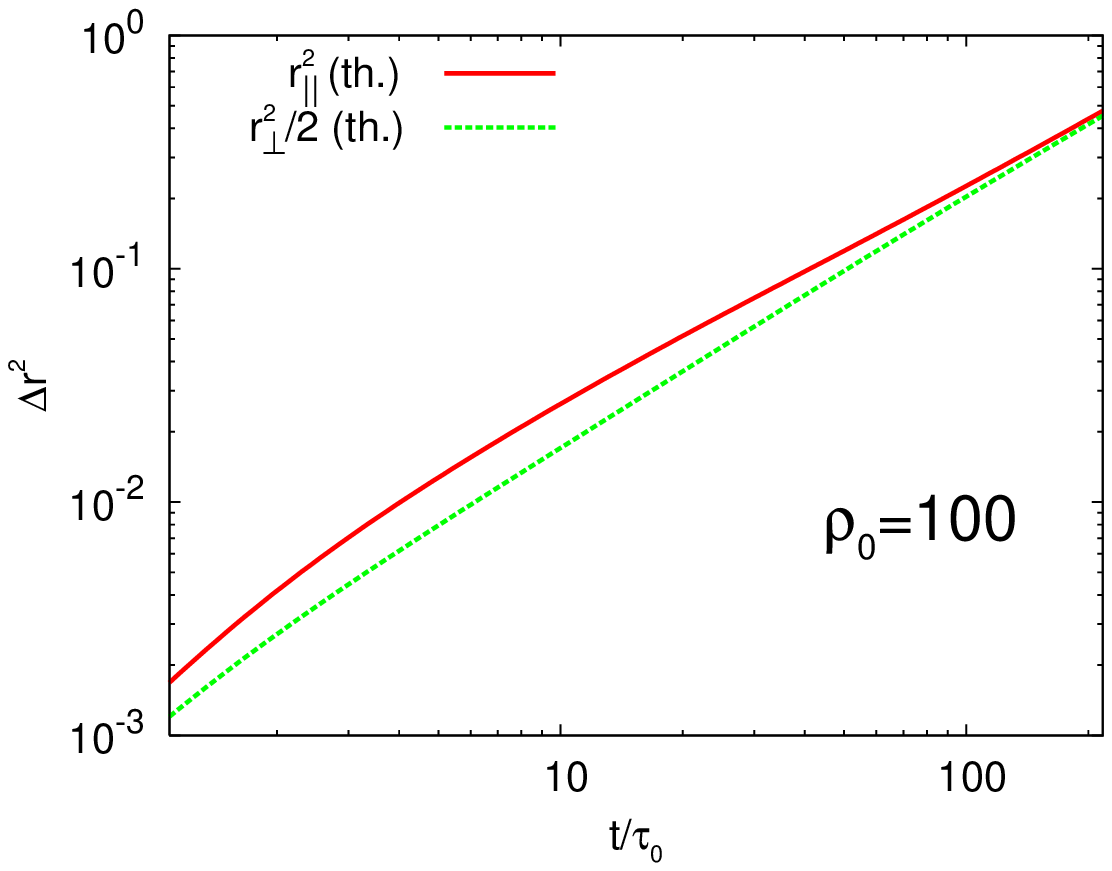}

\vfill
FIG.~\ref{r2figD6b}
\thispagestyle{empty}

\clearpage
\noindent
\includegraphics[width=0.8\textwidth]{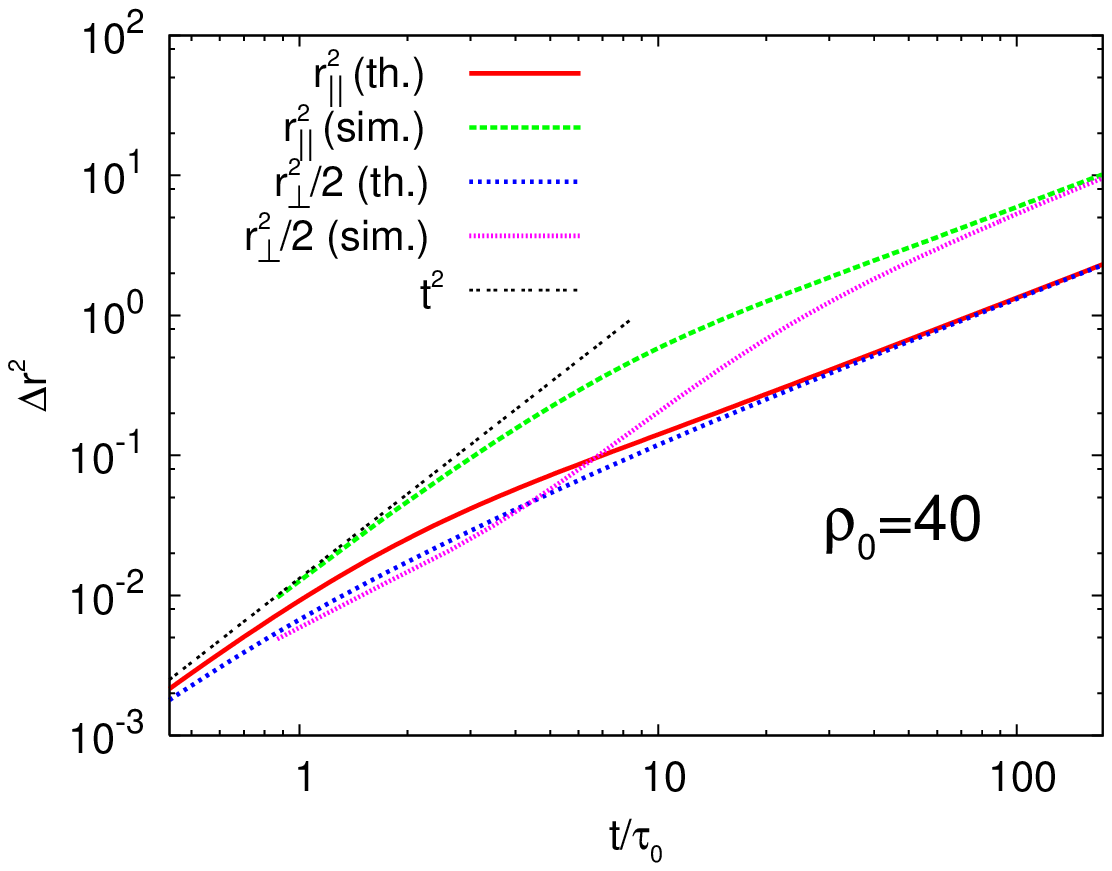}\\
\includegraphics[width=0.8\textwidth]{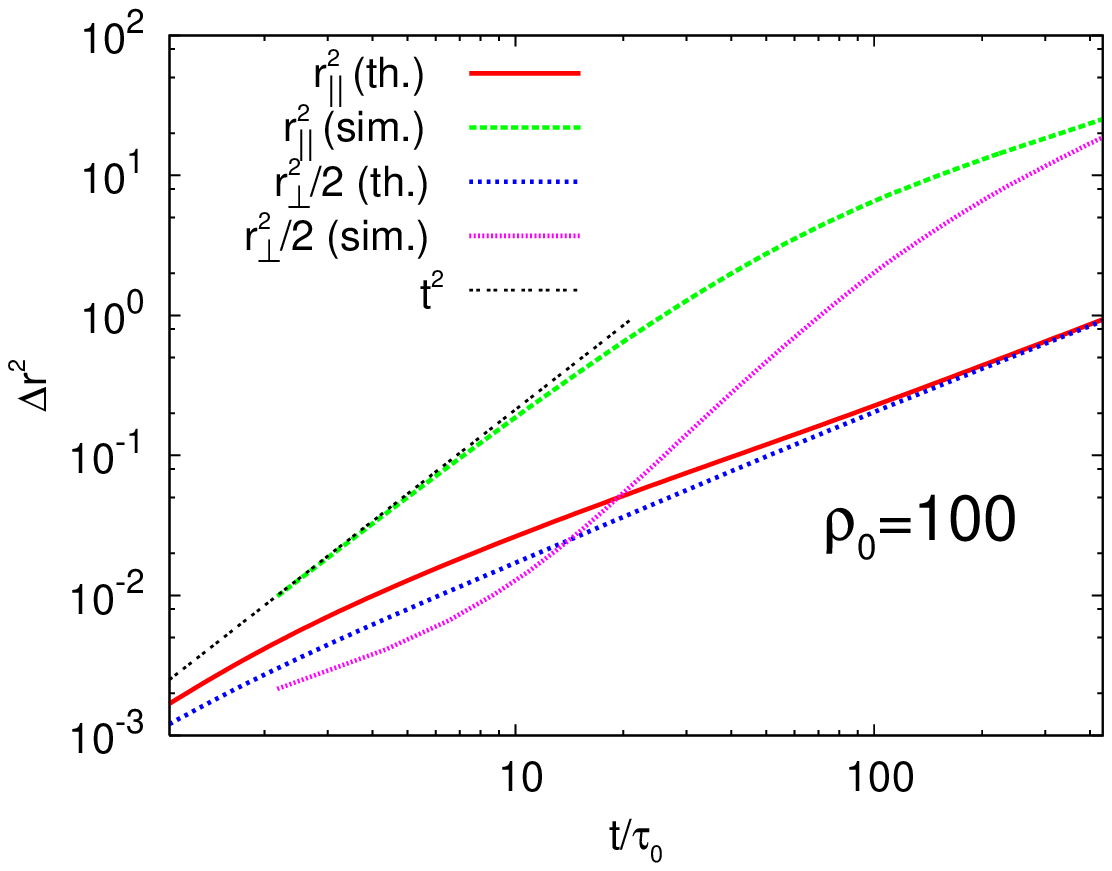}

\vfill
FIG.~\ref{r2figD6c}
\thispagestyle{empty}

\clearpage
\noindent
\includegraphics[width=0.8\textwidth]{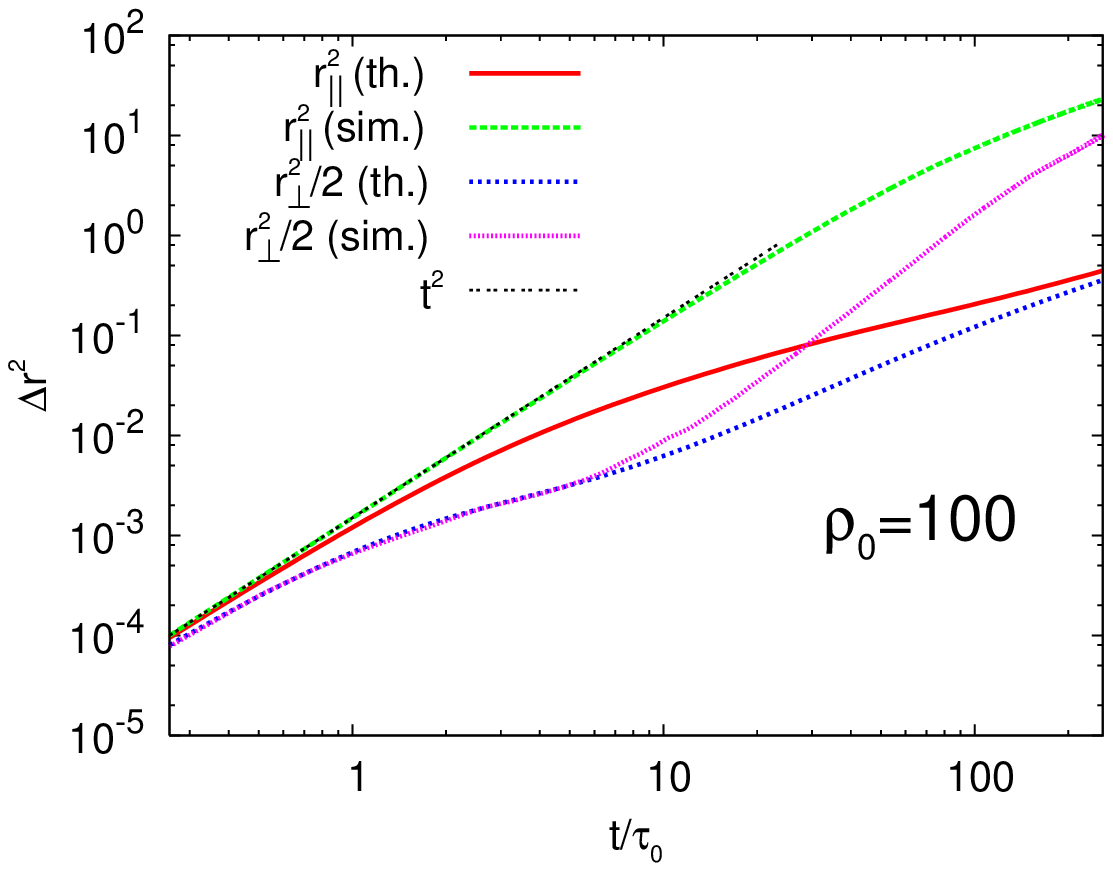}

\vfill
FIG.~\ref{r2figD2}
\thispagestyle{empty}
\end{document}